\documentclass[a4paper,11pt]{article}
\usepackage{jheppub} 
\usepackage{lineno}
\usepackage{lmodern}
\usepackage{lipsum}
\usepackage{etoolbox}
\usepackage{tocloft} 
\usepackage{xpatch}
\usepackage{mathrsfs}
\usepackage{bm}
\usepackage[english]{babel}
\usepackage{amsfonts}
\usepackage{graphicx}
\usepackage[T1]{fontenc}
\usepackage[scr]{rsfso}
\usepackage{mathrsfs}
\usepackage{euscript}
\usepackage{tensor}
\usepackage{stackengine}
\usepackage[dvipsnames]{xcolor}
\usepackage{appendix}
\usepackage[capitalise]{cleveref}
\stackMath

\usepackage[final]{showkeys}

\usepackage{bbm} 
\usepackage{physics}

\usepackage{comment} 

\usepackage{amsmath}
\usepackage{amsthm}
\usepackage{empheq}

\newcommand{\C} {\mathcal{C}}

\usepackage{xcolor}

\usepackage{xcolor}

\usepackage{textgreek}

\newcommand{\beq}{\begin{eqnarray}}
\newcommand{\eeq}{\end{eqnarray}}
\newcommand{\beqn}{\begin{eqnarray}}
\newcommand{\eeqn}{\end{eqnarray}}
\newcommand{\bee}{\begin{equation} \begin{aligned}}
\newcommand{\eee}{ \end{aligned} \end{equation}}
\newcommand{\pa}{\partial}

\newcommand{\cL}{{\cal L}}
\newcommand{\cO}{{\cal O}}

\newcommand{\bF}{{\overline{F}}}
\newcommand{\bR}{{\overline{R}}}
\newcommand{\bP}{J}
\newcommand{\bbeta}{{\overline{\beta}}}
\newcommand{\bU}{{\overline{U}}}

\usepackage{tocloft}
\newcommand{\ve}{\varepsilon}

\newcommand{\cN}{{\cal{N}}}
\newcommand{\cC}{{\cal{C}}}
\newcommand{\cA}{{\cal{A}}}

\newcommand{\bnabla}{{\overline{\nabla}}}

\def\N{{\cal N}}

\newcommand{\chkM}{{\color{red} \,\checkmark\kern-5pt{}_{M}}}

\newcommand{\be}{\begin{equation}}
\newcommand{\ee}{\end{equation}}
\newcommand{\bea}{\begin{eqnarray}}
\newcommand{\eea}{\end{eqnarray}}

\def\pa{\partial}

\newcommand{\scri}{{\mathscr{I}^+}}

\usepackage{amsmath}

\def\l{\lambda}
\def\th{\theta}

\newcommand{\burgos}[1]{
	\centerline{
		\begin{minipage}[c]{0.9\textwidth}
			\begin{center}
		$^a$ Departamento de Física, Universidad de Burgos, 09001 Burgos, Spain
			\end{center}
		\end{minipage}
		}
	}
    
\newcommand{\perimeter}[1]{
	\centerline{
		\begin{minipage}[c]{0.9\textwidth}
			\begin{center}
			$^b$ Physique Th\'eorique et Math\'ematique and International Solvay Institutes\\ Universit\'e Libre de Bruxelles, C.P.231, 1050 Brussels, Belgium
			\end{center}
		\end{minipage}
		}
	}

\newcommand{\chile}[1]{
	\centerline{
		\begin{minipage}[c]{0.9\textwidth}
			\begin{center}
		$^c$ Facultad de Ingeniería, Universidad San Sebastián, Santiago 8420524, Chile
			\end{center}
		\end{minipage}
		}
	}

        \newcommand{\tnt}[1]{
	\centerline{
		\begin{minipage}[c]{1\textwidth}
			\begin{center}
		$^d$ University of Toronto Department of Physics,\\ McLennan Physical Laboratories, 60 St George St, Toronto, ON M5S 1A7, Canada
			\end{center}
		\end{minipage}
		}
	}

    \newcommand{\PI}[1]{
	\centerline{
		\begin{minipage}[c]{0.7\textwidth}
			\begin{center}
			$^e$ Perimeter Institute for Theoretical Physics,\\
			 31 Caroline St. N., Waterloo ON, Canada, N2L 2Y5
			\end{center}
		\end{minipage}
		}
	}

    \newcommand{\chileleo}[1]{
	\centerline{
		\begin{minipage}[c]{0.9\textwidth}
			\begin{center}
		$^f$ School of Physics and Astronomy, Sun Yat-Sen University, Zhuhai 519082, China
			\end{center}
		\end{minipage}
		}
	}
    
\newcommand{\marseille}[1]{
	\centerline{
		\begin{minipage}[c]{0.9\textwidth}
			\begin{center}
		$^g$ Aix Marseille Univ., Univ. de Toulon, CNRS, CPT, UMR 7332,\\ 13288 Marseille, France
			\end{center}
		\end{minipage}
		}
	}

\def\title{BMS Charges from Null Hypersurfaces} 

\begin{document}

{\centering
 \vspace*{1cm}
\textbf{\LARGE{\title{}}}
\vspace{0.3cm}
\begin{center}
Martina Adamo,$^a$ Luca Ciambelli,$^b$ Felipe Diaz,$^c$\\ Michael Imseis,$^{d,e}$ Leonardo Sanhueza,$^f$ Simone Speziale$^{g}$\\
\vspace{-0.2cm}
\textit{\burgos{}}\\
\vspace{-0.1cm}
\textit{\perimeter{}}\\
\vspace{0cm}
\textit{\chile{}}\\
\vspace{-0.1cm}
\textit{\tnt{}}\\
\vspace{0.2cm}
\textit{\PI{}}\\
\vspace{0.2cm}
\textit{\chileleo{}}\\
\vspace{0.2cm}
\textit{\marseille{}}
\end{center}
\vspace{0.6cm}}
\begin{abstract}

We show how the gravitational dynamics, phase space and Noether charges of the BMS group at future null infinity $\scri$ can be derived as limits of their analogues on null hypersurfaces at finite distance. The Bondi mass-loss and angular-momentum evolution equations arise as limits of the Raychaudhuri and Damour constraints. The Ashtekar--Streubel (aka Wald--Zoupas) symplectic potential arises as a specific member of a two-parameter family of null-boundary symplectic potentials, corresponding to the Dirichlet (aka canonical) polarization, while the other members differ by field-space exact terms, leading to the same radiative pre-symplectic structure but potentially different BMS fluxes. 
Deriving the BMS charges requires additional care, since BMS generators extended into the bulk are generally not tangent to the null hypersurfaces. After accounting for this and the associated anomalies, the renormalized canonical charges still contain an additional soft contribution. The prescription of a covariant boundary Lagrangian determines a corner improvement that removes this term and recovers the standard BMS charges.
\end{abstract}

\vspace{0.1cm}
\begin{center}{\small{\href{mailto:marti.adamo@gmail.com}{marti.adamo@gmail.com}, \ \href{mailto:ciambelli.luca@gmail.com}{ciambelli.luca@gmail.com}, \ \href{mailto:felipe.diaz@uss.cl}{felipe.diaz@uss.cl}}}\\
{\small{\href{mailto:michael.imseis@mail.utoronto.ca}{michael.imseis@mail.utoronto.ca}, \ \href{mailto:leonardosanhuezamardones@gmail.com}{leonardosanhuezamardones@gmail.com}, \ \href{mailto:simone.speziale@cpt.univ-mrs.fr}{simone.speziale@cpt.univ-mrs.fr }}}
\end{center}
\thispagestyle{empty}

\newpage
\tableofcontents
\thispagestyle{empty}
\newpage
\clearpage
\pagenumbering{arabic} 

\section{Introduction}

Null hypersurfaces describe black-hole horizons and causal boundaries, and provide the natural arena for the asymptotic description of gravitational radiation from isolated systems~\cite{Penrose:1962ij,Bondi:1962px,Sachs:1962wk}, with a natural common description \cite{Ashtekar:2024stm}. The null geometry is intrinsically different from that of spacelike or timelike hypersurfaces: the induced metric is degenerate, it is possible to identify constraint-free data \cite{Sachs:1962zza}, and more in general describe the geometry in terms of expansion, shear, inaffinity and Hájíček connection \cite{hajivcek1973exact}, see \cite{Newman:1961qr,Henneaux:1979vn,Mars:1993mj,Ashtekar:2000sz,Gourgoulhon:2005ng,Ciambelli:2019lap,Donnay:2019jiz}.
In the last few years, this structure has led to renewed investigations of gravitational dynamics and its phase space on null hypersurfaces at finite distance~\cite{Reisenberger:2007ku,Parattu:2015gga,Lehner:2016vdi,Hopfmuller:2016scf,DePaoli:2017sar,Wieland:2017zkf,DePaoli:2018erh,Hopfmuller:2018fni,Chandrasekaran:2018aop,Oliveri:2019gvm,Chandrasekaran:2020wwn,Chandrasekaran:2021hxc,Chandrasekaran:2021vyu,Adami:2021nnf,Odak:2023pga,Chandrasekaran:2023vzb,Ciambelli:2023mir,Ciambelli:2026vxa,Ciambelli:2026xdn}. A specific outcome relevant to us is the identification of a unique set of Noether charges and their fluxes, associated with the symmetries of generic null hypersurfaces \cite{Chandrasekaran:2018aop}; and more in general of a 2-parameter family obtained relaxing some of the boundary conditions \cite{Odak:2023pga,Chandrasekaran:2023vzb}.
It has also become common to understand the geometry of null hypersurfaces from the intrinsic perspective of Carrollian  geometries \cite{Henneaux:1979vn, Duval:2014uoa}, see also the review~\cite{Ciambelli:2025unn}. 
In parallel, null infinity $\mathscr I^+$ has also received renewed interest, thanks in particular to its role in the physics of gravitational waves (e.g. \cite{Ashtekar:2019viz,Mitman:2021xkq}) and in Strominger's infrared triangles (e.g. \cite{Strominger:2014pwa,Strominger:2017zoo}).  The gravitational data are the news tensor and the asymptotic shear, which describe the Ashtekar--Streubel radiative phase space, and the Bondi energy and angular-momentum aspects, which describe Coulombic data, and are in turn related to the Bondi-van der Burg-Metzner-Sachs  (BMS) symmetries~\cite{Bondi:1962px,Sachs:1962wk,Sachs:1962zza,Geroch:1977big,Ashtekar:1981bq,Wald:1999wa,Barnich:2010eb,Campiglia:2014yka,Flanagan:2015pxa,Madler:2016xju,Speziale:2025lkm}. A specific outcome relevant to our work is the identification of a unique set of BMS charges and fluxes \cite{Geroch:1977big,Ashtekar:1981bq,Dray:1984rfa}, see also \cite{Wald:1999wa,Barnich:2011mi,Flanagan:2015pxa,Chandrasekaran:2021vyu,Grant:2021sxk,Rignon-Bret:2024gcx,Ashtekar:2024stm}.
The geometry of $\scri$ can also be described in Carrollian terms \cite{Bagchi:2016bcd,Ciambelli:2018wre,Fiorucci:2025twa}, which fosters developments in flat-space holography~\cite{Bagchi:2022emh,Donnay:2022aba,Donnay:2022wvx}, including recent constructions of asymptotic boundary stress tensors and gravitational charges~\cite{Freidel:2024emv,Bhambure:2024ftz,Ciambelli:2025mex,Hartong:2025jpp,Hartong:2026rbr,Hartong:2026bvh}. We will use the Carrollian language in this paper.

One of our motivations is to investigate the relation between charges and fluxes on null hypersurfaces at finite distances and the BMS charges and fluxes at $\scri$. In particular, whether there are charges and fluxes at finite distance that reproduce the BMS ones in the limit to $\scri$. This is relevant to strengthen the relation between finite-distance null hypersurfaces and $\scri$ \cite{Ashtekar:2024mme,Ashtekar:2024bpi,Ashtekar:2024stm,Ciambelli:2025mex,Ruzziconi:2025fct,Agrawal:2025fsv}, to investigate notions of gravitational stress tensors useful to flat-holography \cite{Chandrasekaran:2021hxc,Adami:2021kvx,Bhambure:2024ftz, Ciambelli:2025mex,Hartong:2025jpp,Hartong:2026rbr}, and the relation between symmetries on different boundaries. To investigate this question, we need two tools: first, a family of null hypersurfaces parallel to $\scri$, and second, a map between local diffeomorphisms at the null hypersurfaces and asymptotic symmetries. 

For the first tool, we follow the approach of \cite{Ciambelli:2025mex}, where 
a family of null hypersurfaces ${\cal N}_\lambda$ approaching $\mathscr I^+$ for $\l\to\infty$ was constructed using the solution space in Bondi gauge, and their geometry determined by the asymptotic gravitational data. It was shown in \cite{Ciambelli:2025mex} that the Raychaudhuri equation reproduces in the limit the equation for the Bondi mass aspect, and that the null symplectic potential in the Dirichlet polarization of \cite{Parattu:2015gga,Lehner:2016vdi} reproduces in the limit the Ashtekar--Streubel symplectic potential, also known as Wald--Zoupas. 
This choice of symplectic potential is also called canonical in relation to Dirichlet boundary conditions \cite{Ciambelli:2023mir,Ciambelli:2026xdn}, and corresponds to the $(2,2)$ member of the 2-parameter family of \cite{Odak:2023pga,Chandrasekaran:2023vzb}.
A complementary realization of this correspondence was developed in~\cite{Diaz:2026igh}, where gravitational memory was studied using finite-distance Carrollian screens. 
In this work we continue along this line of research, and complete the analysis of \cite{Ciambelli:2025mex} showing that the Damour equation reproduces the equation for the angular momentum aspect in the asymptotic limit, explaining why the null Dirichlet symplectic potential, which was argued to be an unphysical choice because of its gauge-dependence \cite{Lehner:2016vdi,Chandrasekaran:2018aop,Odak:2023pga}, produces a gauge-independent limit to $\scri$, and demonstrating that it is the only one with this property among the 2-parameter family. A key feature is that in our framework, the metric variations at the null hypersurfaces are functions of the asymptotic data determined by the chosen Bondi gauge. Therefore, one cannot invoke any further gauge choice, and \emph{all} variations at the hypersurface have to be kept. Furthermore, we find that also the corner term that distinguishes the covariant and ADM  symplectic potentials \cite{Burnett:1990tww,Harlow:2019yfa,Freidel:2020xyx} has a non-trivial limit to null infinity, unlike for the limit to spatial infinity \cite{Iyer:1994ys}.

For the second tool, we  make the choice of extending the BMS symmetry vector fields off $\scri$ requiring that they preserve the Bondi coordinates that were used to find the null foliation ${\cal N}_\lambda$. This corresponds to the standard formulation of asymptotic symmetries \cite{Bondi:1962px,Sachs:1962wk,Sachs:1962zza}, and can be also given a coordinate-free description \cite{Tamburino:1966zz}. The resulting diffeomorphisms are not tangent to any leaf of ${\cal N}_\lambda$, and  do not preserve any variation on it. They are thus not symmetries in the sense of \cite{Chandrasekaran:2018aop}, and should be rather understood as general diffeomorphisms of ${\cal N}_\lambda$. We compute their fluxes and charges using the 2-parameter family of symplectic potentials \cite{Odak:2023pga}. In doing so, we extend the analysis of \cite{Odak:2023pga} to include a non-tangential component in the diffeomorphism, on top of the arbitrary tangential diffeomorphisms already considered there. The new term gives a Noether charge that is proportional to the inaffinity of the transverse direction to the null hypersurface.
It can also be understood as a specific realization of the universal corner symmetry algebra of \cite{Ciambelli:2021vnn,Freidel:2021cjp,Ciambelli:2022cfr}.
The non-tangential diffeomorphisms also enter the \emph{anomaly} \cite{Hopfmuller:2018fni,Chandrasekaran:2020wwn,Freidel:2021cjp}, namely the difference between field space and spacetime transformations. 

Having obtained the 2-parameter family of generalized charges in terms of the asymptotic data, we evaluate their limit to $\scri$ and compare with the BMS charges. 
After subtracting a divergent Minkowski reference contribution, we obtain a finite family of charges depending on the polarization parameters. 
Following the prescription of \cite{Odak:2022ndm} to select a covariant boundary Lagrangian, we find that the correct BMS charges arise as improved Noether charges corresponding to the Dirichlet polarization. This result shows that it is possible to derive the BMS charges starting from charges on physical null hypersurfaces parallel to $\scri$ in the bulk, and following a specific unique prescription. We also notice that the BMS charges do not correspond to the limit of the null Brown-York tensor \cite{Chandrasekaran:2021hxc}, which receives corrections from both the anomaly contribution, and the non-tangential contribution to the charge at finite distance. Finally, we find that if one weakens the stationarity and covariance prescription of \cite{Odak:2022ndm} by imposing it on the Noether current as opposed to the symplectic potential, then there are two different choices of symplectic potentials that give the same symplectic structure and BMS charges and fluxes at $\scri$: the standard one corresponding to the (2,2) member, and the conformal one corresponding to the (0,1) member.

Our results thus establish a direct correspondence between finite-distance null geometry and the gravitational dynamics, radiative phase space and charges at null infinity. This correspondence involves both a controlled asymptotic limit of the null geometry and a specific bulk extension of the BMS symmetries. While the radiative presymplectic structure is universal within the phase space considered here, identifying the corresponding gravitational charges requires accounting for non-tangential symmetry generators and retaining the finite contributions of boundary Lagrangians, corner terms and their anomalies. The asymptotic limit therefore preserves information that is not captured by the radiative symplectic structure alone.

The paper is organized as follows. Section~\ref{section:geometric setup} develops the null foliation in the physical Bondi spacetime, its intrinsic Carrollian geometry and the asymptotic limit of the gravitational constraints and null Brown--York stress tensor. In Section~\ref{Sec:GravPS}, we study the asymptotic limit of the $(b,c)$ family of gravitational symplectic potentials and their corner contributions. Section~\ref{sec5} presents the construction, renormalization and covariant improvement of the corresponding BMS charges. We conclude in Section~\ref{conclusions} with a summary of the main results and future directions. Appendix~\ref{APPCCS} presents the complementary analysis of the gravitational constraints and symplectic potentials in the conformally compactified spacetime.

\section{Null Hypersurfaces in Bondi Gauge: Geometry and Dynamics}
\label{section:geometric setup}

In this section, we review the construction of null hypersurfaces employed in~\cite{Ciambelli:2025mex}. This serves both to create a self-contained exposition, and to familiarize the reader with the tools we will need. A null hypersurface is a codimension-1 surface in spacetime such that its normal 1-form has vanishing norm. As a consequence, the contra-variant vector associated to the normal is tangent to the hypersurface, hypersurface orthogonal and geodetic, and a null direction of the induced metric. These properties can be also described with an approach that is more intrinsic to the hypersurface, and this is the language of Carrollian geometry \cite{Henneaux:1979vn,Duval:2014uoa,Ciambelli:2019lap,Freidel:2022bai} (see~\cite{Ciambelli:2025unn} for a review). In this language, 
a null hypersurface $\cN$ is specified by a pair ($q_{ab}, \ell^{a}$) such that $\ell^{a}q_{ab}=0$. Here, $x^a$ are arbitrary 3d coordinates, $q_{ab}$ is a degenerate metric and $\ell^{a}$ is a nowhere vanishing vector field. Denoting a cut of $\cal N$ by ${\cal C}$, one can view $\cal N$ as being a line bundle $\pi:{\cal N}\to {\cal C}$.\footnote{More precisely, any regular interval, namely free of conjugate points.} The Carrollian vector field is then an element of the vertical sub-bundle of $T{\cal N}$, the latter defined as the kernel of the pushforward map $\pi_{*}:T{\cal N}\to T{\cal C}$. That is, $\ell\in \text{Ver}(T\cal N)\equiv {\text{Ker}}(\pi_{*})$. Then, a choice of Ehresmann connection $k=k_{a}dx^{a}$ must be made in order to define the horizontal sub-bundle. This object must be a section of the dual bundle $T^*\cN$, satisfying $i_{\ell} k=\ell^a k_a=-1$.\footnote{The  
minus sign here is chosen so that the bulk rigging vector (defined below) is future pointing.} In this section, we will describe how to induce these geometric objects from an ambient bulk spacetime $({\cal M},g)$, which we take to be an asymptotically flat spacetime in Bondi gauge. The formalism we use to do this is the null rigging construction typical of the Newman-Penrose formalism \cite{Newman:1961qr} (see also~\cite{Mars:1993mj}). We apply this formalism in the physical  spacetime, showing how to take the asymptotic limit of the ruled Carrollian data along a family of null hypersurfaces toward future null infinity, $\mathscr{I}^{+}$. It is also useful to perform these computations in the conformally compactified spacetime, aka unphysical spacetime, in which null infinity is a well-defined null hypersurface. We present the unphysical spacetime analysis in Appendix~\ref{APPCCS}.

\subsection{Bondi Gauge \& Null Rigging}

We begin by recalling the metric of an asymptotically flat spacetime in Bondi gauge. Using coordinates $x^\mu=(u,r,x^A)$, 
the line element reads
\begin{equation} \label{2.1}
    ds^{2} = -2e^{2\beta}du\left(dr+Fdu\right) + g_{AB}\left(dx^{A}-U^{A}du\right)\left(dx^{B}-U^{B}du\right)\,,
\end{equation}
where $\beta,F,g_{AB},U^A$ are arbitrary functions thus far. It is  useful to report the non-vanishing components of the metric
\begin{equation}
    g_{ur}=-e^{2\beta}\,,\qquad g_{uu}=g_{AB}U^AU^B-2e^{2\beta}F\,,\qquad g_{AB}\,,\qquad g_{Au}=-g_{AB}U^B\,,
\end{equation}
and its inverse
\begin{equation}
    g^{ur}=-e^{-2\beta}\,,\qquad g^{rr}=2e^{-2\beta}F\,,\qquad g^{AB}\,, \quad g^{Ar}=-e^{-2\beta}U^A\,.
\end{equation}

The asymptotic boundary lives at $r\to \infty$, around which these functions admit the following asymptotic expansions
\begin{align}
    F &= \bF-\frac{m}{r} + \frac{F_2}{r^2}+\mathcal{O}(r^{-3})\,, \label{2.2} \\
    \beta &= \frac{\bbeta}{r^{2}} + \mathcal{O}(r^{-4})\,, \label{2.3}\\
    U^{A} &= \frac{\bU^{A}}{r^{2}} + \frac{U_{1}^{A}}{r^{3}} + {\cal O}(r^{-4})\,, \label{2.4} \\
    g_{AB} &= r^{2}\gamma_{AB} + rC_{AB} +\frac{1}{4}\gamma_{AB} C_{CD}C^{CD} + {\cal O}(r^{-1})\,. \label{2.5}
\end{align}
In the following, it will be convenient to introduce $\bP^{A}$, which is related to $U_{1}^{A}$ via $U_{1}^{A} = -\frac{2}{3}(\bP^{A} + C_{B}{}^{A}\bU^{B} + \partial^{A}\bbeta)$. 

Once expanded in $r$, all quantities appearing in these fall-off conditions are a priori functions of $x^{a}$. The Coulombic data, $m$ and $\bP_{A}$, are the Bondi mass and angular momentum aspects, respectively. The symmetric, traceless (with respect to $\gamma_{AB}$) tensor $C_{AB}$ encodes the radiative gravitational degrees of freedom. $\gamma_{AB}$ is the leading order part of the induced spatial metric on a cut ${\cal C}$ of ${\cal N}$, and is used to raise and lower angular indices. 

The leading order asymptotic Einstein equations yield the following relations
\begin{align} \label{2.6}
    \bbeta +\frac{1}{32}C_{AB}C^{AB}=0\,, \qquad \bR=4\bF\,,  \qquad \bU^{A} +\frac{1}{2}\bnabla _{B}C^{AB} = 0\,,
\end{align}
with $\bnabla_{A}$ and $\bR $ being the Levi-Civita connection and Ricci scalar associated to the metric $\gamma_{AB}$. One also obtains the condition $\pa_{u}\gamma_{AB}=0$ from the leading order Einstein equations, which we will impose throughout this manuscript.\footnote{One can relax this condition by allowing for a time-dependent conformal factor in the metric of a cut. This can be achieved by allowing for an overleading, ${\cal O}(r)$ term in the fall-off for $F$ in~\eqref{2.2}.} We now review how to embed a finite distance null hypersurface in the spacetime endowed with the metric~\eqref{2.1}, and push its geometric data to ${\mathscr{I}}^{+}$ along a null foliation, following~\cite{Ciambelli:2025mex}.

A null hypersurface $\mathcal{N}$ embedded in a spacetime $(\mathcal{M},g)$ is defined by specifying its normal one-form $n$ such that 
$|n|^{2} =0$. We will take it to be
\begin{equation} \label{normalphysical}
    n=-d\left(r-r_{\mathcal{N}}(x^{a})\right)\,,
\end{equation}
so that $\mathcal{N}$ is located at the set of points where $r=r_{\mathcal{N}}(x^{a})$. In other words, we have an embedding $\phi: {\cal N}\to{\cal M}$ such that $x^{a}\to(r_{\cal N}(x^{a}),x^{a})$.
It is important to emphasize that the function $r_{\mathcal{N}}(x^{a})$ must be non-trivial: a hypersurface defined by $r={const}$ in~\eqref{2.1} is timelike and taking $r\to\infty$ specifies a timelike foliation instead of a null one.

The vector field dual to $n$ is readily found to be
\begin{align}\label{nvec}
    n^{\mu}\pa_{\mu} = -e^{-2\beta}\left(2F + (\pa_{u}+U^{A}\pa_{A})r_{\cal N}\right)\pa_{r} + e^{-2\beta}\pa_{u} + \left(g^{AB}\pa_{B}r_{\cal N}+e^{-2\beta}U^{A}\right)\pa_{A}\,,
\end{align}
and we therefore have
\begin{align} \label{n^2}
    |n|^{2} = n^{\mu}n_{\mu} = 2e^{-2\beta}(F+(\pa_{u} + U^{A}\pa_{A})r_{\cal N}) + g^{AB}\pa_{A}r_{\cal N}\pa_{B}r_{\cal N}\,.
\end{align}
We will impose this to vanish perturbatively around the asymptotic boundary.

In addition to the normal one-form $n$, we introduce an auxiliary null vector field, called a null rigging vector, 
\begin{align} \label{rigphys}
    k^{\mu}\pa_{\mu}=\pa_{r}\,,
    \quad \Rightarrow \quad
    k_{\mu}dx^{\mu} = -e^{2\beta}du\,.
\end{align}
By design, the null rigging satisfies $k^{\mu}k_{\mu}=0$ and $k^{\mu}n_{\mu}=-1$.

With~\eqref{normalphysical} and~\eqref{rigphys} in hand, the rigging projector is defined as
\begin{equation}
    \Pi_{\mu}{}^{\nu}=\delta_{\mu}^{\nu}+n_{\mu}k^{\nu}, \qquad \Pi_{\mu}{}^{\sigma}\Pi_{\sigma}{}^{\nu} = \Pi_{\mu}{}^{\nu}\,,
\end{equation}
and projects bulk tensors onto ${\cal N}$, $\Pi_{\mu}{}^{\nu}n_{\nu} = 0$ and $k^{\mu}\Pi_{\mu}{}^{\nu}=0$. Therefore, it can be used to induce the Carrollian geometric data on $\mathcal{N}$ from the bulk\footnote{We will henceforth use the notation $\overset{\cal N}{=}$ to denote equalities holding only on ${\cal N}$.} 
\begin{equation} \label{carrollvec}
    \ell^{a}\overset{\cal N}{=}n^{\mu}\Pi_{\mu}{}^{a}\,,\qquad 
    q_{ab}\overset{\cal N}{=}\Pi_{a}{}^{\mu}\Pi_{b}{}^{\nu}g_{\mu\nu}\,,\qquad 
    k_{a}\overset{\cal N}{=}\Pi_{a}{}^{\mu}k_{\mu}\,.
\end{equation}
By construction, we have that
$\ell^{a}q_{ab}=0$ and $\ell^{a}k_{a}=-1$.

Furthermore, two key geometric quantities on $\cN$ are induced from the bulk Levi-Civita connection. The first is the inaffinity 
\begin{equation}\label{kappa}
    \kappa\overset{\cal N}{=}-k_{\nu}n^{\mu}\nabla_{\mu}n^{\nu}\,,
\end{equation}
which measures the failure of the null generators to be affinely parametrized in the bulk. The second is the Hájíček one-form 
\begin{equation}\label{haj}
  \pi_{a}\overset{\cal N}{=}-q_{a}{}^{\nu}k_{\mu}\nabla_{\nu}n^{\mu}\,,
\end{equation}
where the mixed projector is defined as $q_{a}{}^{\nu}=\delta_{a}{}^{\nu}+k^{\nu}n_{a}+k_{a}n^{\nu}$ and projects bulk objects onto cuts ${\cal C}$. With appropriate assumptions, the Hájíček one-form captures the intrinsic angular momentum along the null hypersurface. These objects enter in the so-called standard Carrollian connection~\cite{Ciambelli:2025unn}.

\subsection{Asymptotic Limit of Null Foliation}
 
We want to solve the eikonal equation for \eqref{normalphysical} in order to find a null foliation of which $\scri$ is one of the leaves, in the unphysical conformal completion. We label the foliation with a parameter $\l$, and require  that the
leaves $\mathcal N_\lambda$
approach future null infinity as $\lambda\to\infty$. Each leaf is
specified by an embedding $r=r_\lambda(x^a)$, and has coordinates $x^a=(u,x^A)$. In the following, we will generally denote a representative
leaf simply by $\mathcal N$, restoring the label $\lambda$
when referring explicitly to the foliation or its
asymptotic limit. We determine the embedding
perturbatively through the inverse-power expansion
\begin{equation} \label{lambdaexp}
    r_{\lambda}(x^a)=\lambda+c_{0}(x^{a})+\frac{c_{1}(x^{a})}{\lambda}+\frac{c_{2}(x^{a})}{\lambda^{2}}+\mathcal{O}(\lambda^{-3})\,.
\end{equation}
The unit coefficient in front of the first term guarantees that $\scri$ will still be described by a Bondi frame. The second term introduces a change in the first order extension of the conformal factor, in other words, a non-zero trace in the next-to-leading order of the angular metric.
Since null hypersurfaces can develop caustics, we
restrict our analysis to smooth portions of the
foliation in the asymptotic region, where the
expansion in inverse powers of $\lambda$ is valid.

To impose the nullity condition, we expand~\eqref{n^2} in $\lambda$, using the asymptotic fall-off conditions~\eqref{2.2}-\eqref{2.5}, and set it to zero order by order. This leads to constraints on the coefficient functions, $c_{i}(x^{a})$, needed to keep the leaves ${\cal N}_\lambda$ null as we approach the boundary. The lowest orders of this procedure give
\begin{equation} \label{coeffconstraints}
   \partial_{u}c_{0}=-\bF \,,\qquad\partial_{u}c_{1}=m\,,\qquad \partial_u c_2 =-c_0 m - F_2\,,
\end{equation}
where, for unit round spheres, one has $\bF=1/2$.
These equations can be immediately integrated, up to angle-dependent arbitrary functions. These represent the left over freedom of choosing different null foliations, and have been used to fix $\pa_A c_0=0$.

The cut curvature and Bondi mass aspect determine the null hypersurfaces.
To make the geometric interpretation of this construction
more explicit, let us consider Schwarzschild spacetime in
retarded Bondi coordinates. Working in the exterior region $r> 2M$, the only nontrivial Bondi
datum is the constant mass aspect $M$, and the relevant
metric functions are
\begin{equation}
    F=\frac12-\frac{M}{r},
    \qquad
    g_{AB}=r^2\gamma_{AB}.
\end{equation}
For a spherically symmetric null hypersurface located
at $r=r_\lambda(u)$, the nullity condition reduces to
\begin{equation}
    \partial_u r_\lambda=-\frac12+\frac{M}{r_\lambda}.
\end{equation}
Choosing the integration constants such that
$r_\lambda(0)=\lambda$, the exact solution is
\begin{equation}
    u=2(\lambda-r_\lambda)
    +4M\ln\left(\frac{\lambda-2M}{r_\lambda-2M}\right),
\end{equation}
which admits the asymptotic expansion
\begin{equation}
    r_\lambda(u)=\lambda-\frac{u}{2}
    +\frac{Mu}{\lambda}
    +\frac{Mu^2}{4\lambda^2}
    +O(\lambda^{-3}),
\end{equation}
in agreement with the general nullity conditions.

The condition $r_\lambda(0)=\lambda$ gives a simple
geometric meaning to the foliation parameter: $\lambda$
is the areal radius at which each null hypersurface
intersects the reference cut $u=0$. These are precisely
the ingoing null hypersurfaces of constant advanced
time $v=u+2r_*(r)$, where
$r_*=r+2M\ln(r/2M-1)$ is the tortoise coordinate.
Evaluating $v$ at the reference cut, we find
\begin{equation}
    v=2r_*(\lambda)
    =2\lambda+4M\ln\left(\frac{\lambda}{2M}-1\right)
    =2\lambda+4M\ln\frac{\lambda}{2M}
    +O(\lambda^{-1}).
\end{equation}
Thus, $\lambda$ coincides with $v/2$ in Minkowski
spacetime, but differs from it by a mass-dependent
logarithmic correction in Schwarzschild. Using $v/2$
directly as the foliation parameter would introduce
logarithms into the asymptotic expansion of the
embedding, whereas our choice preserves a simple
expansion in inverse powers of $\lambda$.

This construction also admits a double-null
interpretation. Since $\partial_\lambda r_\lambda\neq0$
in the exterior region, we may use $\lambda$ as a
coordinate instead of $r$, obtaining
\begin{equation}
    ds^2=-2\partial_\lambda r_\lambda\,du\,d\lambda
    +r_\lambda^2\gamma_{AB}dx^A dx^B.
\end{equation}
The parameter $\lambda$ therefore labels the null
hypersurfaces, while $u$ parametrizes their generators.
At fixed $u$, taking $\lambda\to\infty$ approaches
$\mathscr I^+$, which can be regarded as the limiting
leaf of the foliation in the conformal completion.

The leading embedding reproduces the Minkowski
ingoing foliation, while the subleading terms encode
the effect of the Schwarzschild mass. Although the
shear and H\'aj\'{\i}\v{c}ek one-form vanish in this
example, the null hypersurfaces retain nontrivial
expansion and inaffinity. We now return to general
Bondi data and study the asymptotic limit of these
intrinsic geometric quantities.

\subsection{Limit of Intrinsic Carrollian Data}\label{asym}

We may now expand all our intrinsic Carrollian quantities, as well as the inaffinity and Hájíček one-form, perturbatively in $\lambda$. Let us start with the Carrollian vector field $\ell^{a}$. By construction, since $k^{u}=k^{A}=0$, $\Pi_{\mu}{}^{a}=\delta_{\mu}{}^{a}$. Thus,\footnote{We could have retained the notation $n^a$, but adopt $\ell$ for consistency with the Carrollian conventions used throughout the manuscript, at the price of changing  the usual NP notation in the asymptotic literature, where $l$ is the null vector tangent to the $u$ foliation, here called $k$ instead.}
\begin{align}
    \ell^{a}\overset{\cal N}{=}n^{\mu}\Pi_{\mu}{}^{a}= n^{a} \,.
\end{align}
Therefore, from the vector field~\eqref{nvec} we find
\begin{align}
    \ell^{u} = 1-\frac{2\bbeta}{\lambda^{2}} + {\cal O}(\lambda^{-3})\,,\qquad 
    \ell^{A} =\frac{1}{\lambda^{2}}\bU^{A} + \frac{1}{\lambda^{3}}(U^{A}_{1} + \pa^{A}c_{1} - 2c_{0}\bU^{A}) + {\cal O}(\lambda^{-4})\,.
\end{align}
Next, the components of the induced Carrollian metric are 
\begin{align}
    q_{uu} &= \frac{\bU^A\gamma_{AB}\bU^B}{\lambda^{2}}+ {\cal O}(\lambda^{-3})\,, \\
    q_{uA} &= -\bU_{A} - \frac{1}{\lambda}\left(U_{1\,A}+C_{AB}\bU^{B}+\pa_{A}c_{1}\right) + {\cal O}(\lambda^{-2})\,, \\
    q_{AB} &= \lambda^{2}\gamma_{AB} + \lambda\left(C_{AB}+2c_{0}\gamma_{AB}\right) + q^{(0)}_{AB}+{\cal O}(\lambda^{-1})\,,\label{qAB}
\end{align}
where $q^{(0)}_{AB}=c_0 C_{AB}+( c_0^2+2c_1+\tfrac{1}{4}C_{CD}C^{CD} )\gamma_{AB}$.
The expansion for the Carrollian Ehresmann connection is
\begin{align}
    k_{u} = -1-\frac{2\bbeta}{\lambda^{2}} + {\cal O}(\lambda^{-3})\,, \qquad k_{A}=0\,.
\end{align}
Moreover, from the definitions~\eqref{kappa},~\eqref{haj}, we can expand the inaffinity and Hájíček one-form
\begin{align}
    \kappa &= -\frac{m}{\lambda^{2}} + {\cal O}(\lambda^{-3})\,, \label{kexp}\\
    \pi_{u}&= {\cal O}(\lambda^{-3})\,,
    \qquad \pi_{A} = \frac{1}{\lambda}\bU_A  {}-\frac{1}{\lambda^2}\left(\bP_{A} + 2\pa_{A}\bbeta +  c_{0}\bU_A   - \partial_Ac_1\right) + {\cal O}(\lambda^{-3})\,. \label{Hajphys}
\end{align}

With the above expansions in $\lambda$, let us now introduce a few more quantities that are relevant for our analysis. The first is the expansion tensor, or second fundamental form,
\begin{align} \label{exptensor}
    \theta_{ab} = \frac{1}{2}\mathcal{L}_{\ell}q_{ab}\,,
\end{align}
defined via data which are purely intrinsic to ${\cal N}$, satisfying $\ell^{a}\theta_{ab}=0$. Its components can be expanded asymptotically, yielding
\begin{align}
    \theta_{uu} &= {\cal O}(\lambda^{-2})\,, \\
    \theta_{uA} &= -\frac{1}{\lambda}\bU^{B}\left(\frac{1}{2}N_{AB} - \bF \gamma_{AB}\right) + {\cal O}(\lambda^{-2})\,, \\
    \theta_{AB} &= \lambda\left(\frac{1}{2}N_{AB} - \bF \gamma_{AB}\right) + \frac{1}{2}\mathcal{L}_{\bU}\gamma_{AB}-\frac12 \bF C_{AB}+\frac12 c_0 N_{AB}\notag\\
    &\quad +\gamma_{AB}\left(m+\frac14 C_{CD}N^{CD}-c_0 \bF\right)+ {\cal O}(\lambda^{-1})\,,
\end{align}
with $N_{AB}=\pa_{u}C_{AB}$ the News tensor.

The mixed-indices tensor $\theta_{a}{}^{b}$, defined via
$\theta_{ab} = \theta_{a}{}^{c}q_{cb}$ and $\theta_{a}{}^{b}k_{b}=0$, can be decomposed into its trace and trace-free parts 
\begin{align}
    \theta_{a}{}^{b} = \frac{1}{2}\theta q_{a}{}^{b} + \sigma_{a}{}^{b}\,,
\end{align}
where $\theta$ is the expansion scalar, $q_{a}{}^{b}=\delta_{a}{}^{b}+k_{a}\ell^{b}$ is the horizontal projector, and $\sigma_{a}{}^{b}$ is the traceless shear. The second condition in defining $\theta_{a}{}^{b}$ implies $\theta_{a}{}^{u}=0$. One can then use the first condition to find the following expansions
\begin{align}
    \theta_{u}{}^{A} &= {{\cal O}(\lambda^{-3})}\,, \\
    \theta_{A}{}^{B} &= \frac{1}{\lambda}\left(\frac{1}{2}N_{A}{}^{B}-\delta_{A}{}^{B}\bF \right) + \frac{1}{\lambda^{2}}\Big[\delta_{A}{}^{B}\left(c_{0}\bF +m+\frac{1}{4}N_{CD}C^{CD}\right)\notag 
    \\ &  \quad + \frac{1}{2}\left(\bF C_{A}{}^{B} - N_{AC}C^{CB}-c_{0}N_{A}{}^{B} + 2\bnabla_{(A} \bU^{B)} \right)\Big] + {\cal O}(\lambda^{-3})\,, 
\end{align}
where $\bnabla_{(A} \bU^{B)}$ denotes the symmetric part of $\bnabla_{A} \bU^{B}$.
Now, taking the trace yields the expansion scalar
\begin{align}
    \theta = -\frac{2}{\lambda}\bF  + \frac{1}{\lambda^{2}}\left(2c_{0}\bF  + 2m + \bnabla _{A}\bU^{A}\right) + {\cal O}(\lambda^{-3})\,,\label{thexp}
\end{align}
while the components of the shear are
\begin{align}
    \sigma_{a}{}^{u} &= 0\,, \label{physshear}\qquad \qquad 
    \sigma_{u}{}^{A} = {\cal O}(\lambda^{-3})\,, \\
    \sigma_{A}{}^{B} &= \frac{1}{2\lambda}N_{A}{}^{B} + \frac{1}{4\lambda^{2}}\left(C_{AC}N^{BC} - N_{AC}C^{BC} + 2(\bF C_{A}{}^{B} - c_{0}N_{A}{}^{B} + 2\bnabla_{\langle{A}}\bU^{B\rangle})\right) + {\cal O}(\lambda^{-3})\,,\nonumber
\end{align}
where we used the identity $\delta_A^B N_{CD}C^{CD}=C_{AC}N^{BC}+N_{AC}C^{CB}$, and $\bnabla_{\langle{A}}\bU^{B\rangle}$ denotes the symmetric and traceless part of $\bnabla_{A} \bU^{B}$.

This concludes our review of the asymptotic limit of the intrinsic Carrollian data in the physical Bondi spacetime. These expansions will be central for analyzing the asymptotic limit of the intrinsic gravitational constraints on ${\cal N}$ (the Raychaudhuri and Damour constraints) and for the novel part of this paper, the analysis of the symplectic potential and charges.

\subsection{Asymptotic Einstein Equations}\label{sec3}

The gravitational constraints induced on null hypersurfaces are the Raychaudhuri $C$ and Damour $C_a$ constraints
\begin{align}
    C &= (\mathcal{L}_{\ell}+\theta)\theta -\mu\theta+\sigma_{a}{}^{b}\sigma_{b}{}^{a}+\ell^a\ell^b R_{ab}=0\,, \label{Rconstraint} \\
    C_{a} &= (\mathcal{L}_{\ell}+\theta)\pi_{a}+\theta\varphi_{a} - (\bar{D}_{b}+\varphi_{b})(\mu q_{a}{}^{b} - \sigma_{a}{}^{b})- q_a{}^b R_{bc}\ell^c=0\,,\label{Dconstraint}
\end{align}
where $\mu=\kappa+\frac{\theta}{2}$, and $\varphi_{a}=\mathcal{L}_{\ell}k_{a}$ is the Carrollian acceleration. Together, these equations describe the gravitational dynamics intrinsic to null hypersurfaces.

These equations and the manipulations that follow can be written without any reference to the Carrollian language, but we chose to do so to highlight this perspective. Then, $D_a$ is the Carrollian connection. Given
a tensor $\cA_a{}^b$, the horizontally-projected covariant derivative is given by $\bar D_b \cA_a{}^b=q_b{}^c q_a{}^d q_e{}^b D_c \cA_d{}^e=q_e{}^c q_a{}^d D_c \cA_d{}^e$. The standard Carrollian connection is non-metrical but torsionless. Given an ambient manifold, as in the present case, it is given by projecting on $\cN$ the bulk Levi-Civita connection, using the Rigging projector. As was shown in~\cite{Ciambelli:2025unn}, the affine symbols associated to $D_a$ are given by
\beq\label{standard}
    \Gamma_{bc}^{a}=\left(-\left(\partial_{b}+\omega_{b}\right)k_{c}-k_{b}\left(\pi_{c}+\varphi_{c}\right)-\bar{\theta}_{bc}\right)\ell^{a}+\frac{q^{ad}}{2}\left(\partial_{b}q_{cd}+\partial_{c}q_{bd}-\partial_{d}q_{bc}\right)\,,
\eeq
where $\omega_a=-\kappa k_a+\pi_a$ and $\bar\theta_{bc}=-q_b{}^{\mu}q_c{}^{\nu}\nabla_\mu k_\nu$. Most of the tensors entering~\eqref{standard} will be subleading in the asymptotic limit. 

We now study the asymptotic limit of the constraints~\eqref{Rconstraint} and~\eqref{Dconstraint} along the null foliation constructed
in the previous section. The asymptotic limit of the
Raychaudhuri constraint was shown in \cite{Ciambelli:2025mex} to
reproduce the Bondi mass equation.
Here, we extend
this correspondence to the Damour constraint, showing
that its asymptotic limit reproduces the angular-momentum
evolution equation. As we will see, the leading
contributions cancel, and the Bondi evolution equations
emerge at order $\lambda^{-2}$ in the expansion of
the intrinsic null constraints.

Consider first the Raychaudhuri constraint~\eqref{Rconstraint}, and expand this constraint up to ${\cal O}(\lambda^{-2})$. The first building block is
\begin{align}
(\mathcal{L}_{\ell}+\theta)\theta
    &= \frac{1}{\lambda^{2}}\left(2\bF ^{2}+2\pa_{u}m +\bnabla _{A}\pa_{u}\bU^{A}\right) +{\cal O}(\lambda^{-3})\,.
\end{align}
Next, we have
\begin{align}
    \mu\theta = \frac{2\bF ^{2}}{\lambda^{2}} + {\cal O}(\lambda^{-3})\,,\qquad \sigma_{a}{}^{b}\sigma_{b}{}^{a} = \frac{1}{4\lambda^{2}}N_{AB}N^{AB} + {\cal O}(\lambda^{-3})\,.
\end{align}
Assuming that we are in vacuum, so that $\ell^a\ell^b R_{ab}=0$, and putting everything together, we have obtained
\begin{align}
    C = \frac{1}{\lambda^{2}}\left(2\pa_{u}m -\frac{1}{2}\bnabla _{A}\bnabla _{B}N^{AB}  + \frac{1}{4}N_{AB}N^{AB}\right) + {\cal O}(\lambda^{-3})\,,
\end{align}
where we have used the on-shell relation~\eqref{2.6}. Therefore, going on-shell of the Raychaudhuri constraint gives
\begin{align}\label{BMLF}
    \pa_{u}m = \frac{1}{4}\bnabla _{A}\bnabla _{B}N^{AB}  - \frac{1}{8}N_{AB}N^{AB}\,, 
\end{align}
which is exactly the Bondi mass loss formula.

We now turn to the Damour constraint. Using~\eqref{Hajphys} and~\eqref{physshear}, the temporal component of~\eqref{Dconstraint} can be shown to be subleading, $C_u=\cO(\lambda^{-3})$.
Thus, we focus on $C_{A}$. First, let us write the asymptotic limit of the Carrollian acceleration. From the form of the Ehresmann connection, we find
\begin{align}
    \varphi_{u} =\cO(\lambda^{-4})\,, \qquad \varphi_{A} = \frac{2}{\lambda^{2}}\pa_{A}\bbeta + {\cal O}(\lambda^{-3})\,.
\end{align}
Given that $\varphi_A$ in~\eqref{Dconstraint} multiplies quantities that start at best at order $\frac1{\lambda}$, we thus have that every term in the Damour constraint containing $\varphi_{A}$ will be ${\cal O}(\lambda^{-3})$ and can be neglected. Next, since $q_A{}^u=\delta_A{}^u+k_A \ell^u=0$, $q_u{}^A=k_u\ell^A$, and $q_A{}^B=\delta_A{}^B$, one has
\beq
\bar D_b (\mu q_A{}^b)= \pa_A \mu\,,\qquad \cL_\ell \pi_A= \pa_{u}\pi_{A} + {\cal O}(\lambda^{-3})\,.
\eeq
Then, using that in vacuum $q_a{}^b R_{bc}\ell^c=0$,  the equation we need to analyze is
\begin{align} \label{DamourA}
    C_{A}= \pa_{u}\pi_{A}+\theta\pi_{A} - \pa_{A}\mu + \bar D_b\sigma_A{}^b+ {\cal O}(\lambda^{-3})\,.
\end{align}

From the expansion of the Hájíček one-form in~\eqref{Hajphys}, the first term is
\beq \label{pa_u pi_a}
    \pa_{u}\pi_{A} = -\frac{\bnabla_{B}N_{A}{}^{B}}{2\lambda}+ \frac{1}{\lambda^{2}}\Bigg(\bnabla_{A}\Big(\frac{1}{8}N_{BC}C^{BC}+m\Big)-\pa_{u}(\bP_{A}+c_0 \bU_{A}) \Bigg) + {\cal O}(\lambda^{-3})\,,
\eeq
where we have used the on-shell relations in~\eqref{2.6}, as well as our nullity constraints~\eqref{coeffconstraints}.
Next, using~\eqref{kexp} and~\eqref{thexp}, we have
\begin{align}
    \theta\pi_{A} = -\frac{2\bF }{\lambda^{2}}\bU_{A} + {\cal O}(\lambda^{-3})\,,\qquad 
    \pa _{A}\mu = \frac{1}{2\lambda^{2}}\bnabla _{A}\bnabla _{B}\bU^{B} + {\cal O}(\lambda^{-3})\,.\label{T3}
\end{align}
Eventually, using the affine symbols~\eqref{standard}, the last term in~\eqref{DamourA} can be processed into
\begin{align} 
    \bar D_b\sigma_A{}^b=\frac{1}{2\lambda}\bnabla _{B}N_{A}{}^{B} +\frac{1}{\lambda^{2}}\left( \bnabla _{B}\overset{(2)}{\sigma}_{A}{}^{B} -\frac{1}{4}N^{BC}\bnabla _{A}C_{BC} \right) + {\cal O}(\lambda^{-3})\,,\label{T4}
\end{align}
where 
\begin{align}
    \overset{(2)}{\sigma}_{A}{}^{B} &= \frac{1}{4}\left(C_{AC}N^{BC} - N_{AC}C^{CB} + 2(\bF C_{A}{}^{B} - c_{0}N_{A}{}^{B} 
    + 2\bnabla_{\langle{A}}\bU^{B\rangle})\right)
\end{align}
is the ${\cal O}(\lambda^{-2})$ term in the expression for the shear in~\eqref{physshear}. In particular, the subleading contribution to the
projected divergence involves not only the
$\lambda^{-2}$ term of the shear, but also the
corrections to the connection induced by the asymptotic
expansion of the cut metric. Both contributions are
needed to recover the complete angular-momentum
evolution equation. The derivation of~\eqref{T4} is  non-trivial. We will not display the details here, since a very similar analysis is carried out in Appendix~\ref{APPCCS}, where we show the detailed derivation.

Adding up all these contributions, namely~\eqref{pa_u pi_a},~\eqref{T3}, and~\eqref{T4}, we first observe that at order $\cO(\lambda^{-1})$, the contributions from $\partial_u\pi_A$ and the projected shear divergence cancel identically, so that the first non-trivial Damour constraint arises at $\cO(\lambda^{-2})$. Then, we arrive at the following asymptotic expansion of the spatial part of the Damour constraint
\begin{align} \label{3.16}
        C_{A} &= \frac{1}{\lambda^{2}}\Big[-\pa_{u}\bP_{A}+\bnabla _A m  +\frac{1}{8} \bnabla _{A}(N_{BC}C^{BC}) -\frac{1}{4}N^{BC}\bnabla _{A}C_{BC} + \frac{1}{4}\bnabla _{C}(\bnabla _{A}\bnabla _{B}C^{BC}\nonumber\\
        &-\bnabla ^{C}\bnabla ^{B}C_{A}{}_{B})
    +  \frac{1}{4}\bnabla _{B}(C_{AC}N^{BC}-N_{AC}C^{BC})+ [\bnabla _{B},\bnabla _{A}]\bU^{B} - 2\bF \,\bU_{A}\Big] + {\cal O}(\lambda^{-3})\,,
\end{align}
where we used the identity
\begin{align}
\bnabla_B\bnabla_{\langle{A}}\bU^{B\rangle}= \frac12\bnabla^2\bU_A +\bF\,\bU_A\,.
\end{align}
Processing the second-to-last term in~\eqref{3.16}, we have
\begin{align}
    [\bnabla _{B},\bnabla _{A}]\bU^{B} = \bR _{AB}\bU^{B}=  2\bF \,\bU_{A}\,, 
\end{align}
and thus the last two terms in~\eqref{3.16} cancel. 

Therefore, going on-shell of the asymptotic Damour constraint yields the result
\begin{align}
    \pa_{u}\bP_{A} &= \bnabla _A m  +\frac{1}{8} \bnabla _{A}(N_{BC}C^{BC}) + \frac{1}{4}\bnabla _{B}\left(C_{AC}N^{BC}-N_{AC}C^{BC}\right)\notag \\
    &\quad+ \frac{1}{4}\bnabla _{C}(\bnabla _{A}\bnabla _{B}C^{BC}-\bnabla ^{C}\bnabla^{B}C_{AB}) -\frac{1}{4}N^{BC}\bnabla _{A}C_{BC}\,.
\end{align}
This reproduces the Bondi angular-momentum evolution
equation. Together with the Raychaudhuri result \eqref{BMLF},
it establishes that both Bondi evolution equations
arise as the asymptotic limit of the gravitational
constraints intrinsic to finite-distance null
hypersurfaces. This completes our analysis of the
asymptotic gravitational dynamics.

\subsection{Null Brown--York Stress Tensor}\label{NBY}

An interesting recent result is the relation between the intrinsic gravitational constraints on ${\cal N}$ and the conservation laws for a null analogue of the Brown--York stress tensor, see e.g. \cite{Hopfmuller:2018fni,Chandrasekaran:2021hxc,Ciambelli:2025unn}. Let us define the null stress tensor
\begin{align} \label{NullBY}
    T_{a}{}^{b} = \tau_{a}{}^{b} + \tau_{a}\ell^{b}=\frac1{8\pi G}(W_a{}^b-W \delta_a{}^b)\,,
\end{align}
where
\begin{align}
    \tau_{a}{}^{b} = \frac{1}{8\pi G}\left(\sigma_{a}{}^{b}-\mu q_{a}{}^{b}\right)\,, \qquad \tau_{a} = \frac{1}{8\pi G}\left(\pi_{a}+\theta k_a\right)\,,
\end{align}
and $W_a{}^b=\Pi_a{}^\mu \nabla_\mu n^\nu \Pi_\nu{}^b$  is the Weingarten map, with $W=\theta+\kappa$ its trace, and $T=-W/(4\pi G)$. 

In vacuum, the gravitational constraints are encoded in the conservation law
\begin{align}
    D_{b}T_{a}{}^{b}=0\,, 
\end{align}
whose temporal and horizontal projections reproduce respectively the Raychaudhuri~\eqref{Rconstraint} and Damour~\eqref{Dconstraint} equations.\footnote{This is the null hypersurface version of the ADM constraint $D_b\Pi_a{}^b=0$.} As these constraints reproduce in the limit to $\scri$ the mass and angular momentum equations, it is natural to ask whether the limit to $\scri$ of the null Brown--York stress tensor~\eqref{NullBY} produces a useful notion of gravitational stress tensor for asymptotically flat spacetimes. 

Using the asymptotic expansions of the previous section, we find
\begin{align}
    T_{u}{}^{u} &= \frac{\bF }{4\pi G\lambda} - \frac{1}{8\pi G\lambda^{2}}\left(2c_{0}\bF +2m+\bnabla _{A}\bU^{A}\right) + {\cal O}(\lambda^{-3})\,, \\ 
    T_{u}{}^{A}&= {\cal O}(\lambda^{-3}), \\
    T_{A}{}^{u} &= \frac{1}{8\pi G\lambda}\bU_{A} - \frac{1}{8\pi G\lambda^{2}}\left(\bP_{A}+2\bnabla _{A}\bbeta-\bnabla _{A}c_{1}+c_{0}\bU_{A}\right) +  {\cal O}(\lambda^{-3})\,, \\
    T_{A}{}^{B} &= \frac{1}{16\pi G\lambda}\left( N_{A}{}^{B} + 2\bF \delta_{A}{}^{B}\right) + \frac{1}{16\pi G\lambda^{2}}\left(2\overset{(2)}{\sigma}_{A}{}^{B}-\left(2c_{0}\bF +\bnabla _{C}\bU^{C}\right)\delta_{A}{}^{B}\right) + {\cal O}(\lambda^{-3})\,.
\end{align}

The first thing to notice is that some of the leading $1/\lambda$ terms  precisely match the limit of the corresponding geometric quantity in Minkowski. We can therefore renormalize the null Brown-York tensor subtracting off the Minkowski contributions. The same situation occurs in the limit of the Hawking energy to null infinity \cite{Ashtekar:2024xwl} and in the limit of the Brown--York charges to spatial infinity \cite{Brown:1992br}. Then, one can in principle construct the Brown--York-like charges by contracting this stress tensor with the asymptotic  vector field generating the BMS symmetry. We will do so later (see \eqref{qbcCarrollGeneral}), but we already anticipate that this does not reproduce the standard  BMS charges, due to anomalies and corner improvements, which can be dealt with following the prescription of \cite{Odak:2022ndm}. 

This provides an extension of the results on the limit of codimension-2 Brown-York stress tensor \cite{Brown:1996bw}, and the answer
can be compared to the similar situation at spatial infinity, where the Brown--York charges reproduce the ADM ones (after a suitable subtraction of divergences) \cite{Brown:1992br}, and to the Balasubramanian--Kraus stress tensor for asymptotically AdS spacetimes~\cite{Balasubramanian:1999re}.
Concerning potential holographic interpretations, we remark that the leading order part of the 2d trace is ${T}_A{}^A=\frac{1}{16\pi G \lambda}\bR+\mathcal{O}(\lambda^{-2})\,.$
Rescaling this trace by a factor of $\lambda$ and formally comparing it with the standard two-dimensional trace-anomaly expression
\begin{equation}
    \langle T_A{}^A\rangle =
    \frac{c}{24\pi}\bR\,,
\end{equation}
one would identify the effective central charge
\begin{equation}
    c_{\rm eff} = \frac{3}{2G}\,.
\end{equation}
Since $G$ is dimensionful in four spacetime dimensions, $c_{\rm eff}$ is not a conventional dimensionless CFT central charge.\footnote{Ultimately, this is due to the fact that flat space has no other scales, whereas in AdS one can use the cosmological constant.} Nevertheless, the resemblance suggests that the asymptotic null Brown--York tensor may provide a useful starting point for investigating trace-anomaly-like structures in celestial holography, echoing related developments in the literature~\cite{Bhambure:2024ftz, Baulieu:2025itt, Fiorucci:2025twa, Hartong:2025jpp}, and so it would be interesting to pursue this study further.

\section{Gravitational Phase Space}\label{Sec:GravPS}

Having reviewed the geometric setup, we are now ready to investigate the asymptotic limit of the gravitational phase space induced on null hypersurfaces. We build upon the phase space construction in~\cite{Odak:2023pga}, placing particular emphasis on the two-parameter family of symplectic potentials that arise on finite distance null hypersurfaces. Within our framework, we then take their asymptotic limit. We relegate to Appendix~\ref{APPCCS} the symplectic analysis in the conformally compactified spacetime.

\subsection{Null Phase Spaces at Finite Distance}
\label{section:null phase spaces}

Let us begin by recalling the Einstein--Hilbert symplectic potential written as a spacetime three-form
\begin{align} \label{EHPot}
    \theta_{\rm EH} = \frac{1}{16\pi G}\theta_{\rm EH}^{\mu}\varepsilon_{\mu\nu\rho\sigma}\frac{1}{3!}dx^{\nu}\wedge dx^{\rho}\wedge dx^{\sigma}\,,\qquad \theta_{\rm EH}^{\mu} = 2g^{\rho[\sigma}\delta\Gamma^{\mu]}_{\rho\sigma}\,.
\end{align}
We denote the pullback to a finite distance null hypersurface ${\cal N}$ by $\leftarrow$. 
For our considerations, it is crucial that we keep the variations completely arbitrary.\footnote{For alternative expressions with some restrictions, see e.g. \cite{Lehner:2016vdi,Hopfmuller:2016scf,Chandrasekaran:2020wwn}.} This gives \cite{Parattu:2015gga,Oliveri:2019gvm}
\begin{align} \label{4.3}
    \underset{\longleftarrow}{\theta_{\rm EH}}=\frac{1}{16\pi G} \left[\left(\sigma^{ab} + \frac{\theta}{2}q^{ab}\right)\delta q_{ab} -2\omega_{a}\delta\ell^{a} + 2\delta(\kappa + \theta)\right]\ve_{\cal N} + d\vartheta^{\rm EH}\,,
\end{align}
where $\ve_{\cal N} := i_k\ve_4 = e^{2\beta}\sqrt{{\rm det} q_{AB}}du\wedge d^2x$ and $\omega_{a} = \pi_{a}-\kappa k_{a}$. The last term in~\eqref{4.3} is the corner term
\begin{align} \label{4.5}
    \vartheta^{\rm EH} = \frac{1}{16\pi G}\left((k^{\mu}\delta n_{\mu} + k_{\mu}\delta n^{\mu})\ve_{\cal C} - k\wedge i_{\delta n}\ve_{\cC}\right)\,,
\end{align}
with $\ve_{\cC} := i_\ell \ve_{\cal N}$, which pulls back to the volume form $\sqrt{{\rm det} q_{AB}}d^2x$ on a cut. 

Using the identity
\begin{align} \label{4.7}
    \delta\ve_{\cal N} = \left[\frac{1}{2}q^{ab}\delta q_{ab} + k_{a}\delta\ell^{a}\right]\ve_{\cal N}\,,
\end{align}
and integrating the third term in~\eqref{4.3} by parts in field space, we can rewrite
\begin{align}
  \underset{\longleftarrow}{\theta_{\rm EH}} &= \frac{1}{16\pi G}\left[(\sigma^{ab}-\mu q^{ab})\delta q_{ab} - 2(\pi_{a}+\theta k_{a})\delta\ell^{a}\right]\ve_{\cal N} + \frac{1}{8\pi G}\delta[(\kappa+\theta)\ve_{\cal N}] + d\vartheta^{\rm EH} \notag \\
    &= \theta^{{\prime}}_{\rm D} - \delta\ell_{\rm D} + d\vartheta^{\rm EH}\,, \label{4.8}
\end{align}
where
\begin{align}\label{ellD}
    \ell_{D} \equiv -\frac{1}{8\pi G}(\kappa + \theta)\ve_{\cal N}.
\end{align}
The subscript D indicates that in this choice of phase space polarization, we are imposing Dirichlet boundary conditions on the data $(q_{ab},\ell^{a})$ \cite{Lehner:2016vdi}. In~\cite{Ciambelli:2023mir}, $\theta^{\prime}_{\rm D}$ was called the canonical potential, $\theta^{\rm can}$, and the boundary conditions correspond to fixing the intrinsic Carrollian data on the null hypersurface. We can also manifestly see that \eqref{ellD}
is the boundary Lagrangian needed to be added to the Einstein--Hilbert action to have a well-defined variational principle when imposing Dirichlet boundary conditions on the Carrollian data.

We can consider other polarizations, or more in general, any symplectic potentials in the equivalence class 
\begin{align} \label{4.6}
    \theta = \theta^{\prime} - \delta\ell + d\vartheta\,.
\end{align}
This decomposition identifies a polarization if  $\theta^{\prime}=p_i\delta q^i$ for some set of independent configuration variables $q^i$. It identifies conservative boundary conditions corresponding to $\delta q^i=0$ for all $i$, so that $\theta^\prime$ vanishes on the allowed variations. 
Changing polarization in the spin-0 sector of the phase space, namely the inaffinity $\kappa$ and expansion $\theta$, leads to a two-parameter family of boundary Lagrangians~\cite{Odak:2023pga} 
\begin{align}
    \ell^{(b,c)} = -\frac{1}{16\pi G}(b\kappa +c\theta)\ve_{\cal N}\,.
\end{align}
This induces on ${\cal N}$ a two-parameter family of symplectic potentials via
\begin{align}\label{EHtheta}
    \underset{\longleftarrow}{\theta_{\rm EH}} = \theta^{(b,c)} - \delta\ell^{(b,c)} + d\vartheta^{\rm EH}\,,
\end{align}
with
\begin{equation}\label{4.13}
    \theta^{(b,c)} = \frac{1}{16\pi G}\left((\sigma^{ab}\delta q_{ab} - \varpi_{a}\delta\ell^{a} + (2-b)\delta\kappa + (2-c)\delta\theta)\ve_{\cal N}
 -(b\kappa+(c-1)\theta)\delta\ve_{\cal N}\right)\,,
\end{equation} 
where we have introduced $\varpi_{a} = 2(\pi_{a} -(\kappa - \frac{\theta}{2})k_{a})$. Again making use of the identity in~\eqref{4.7}, one can rewrite $\theta^{(b,c)}$ in the convenient form
\begin{align} \label{thetabc}
    \theta^{(b,c)} = \theta^{\rm can} +\frac{(2-b)}{16\pi G}\delta(\kappa\ve_{\cal N}) + \frac{(2-c)}{16\pi G}\delta(\theta\ve_{\cal N})\,.
\end{align}
Written in this way, it is easy to see that the choice $(b,c)=(2,2)$ yields the Dirichlet or canonical polarization~\eqref{4.8}, while the choice $(b,c)=(0,0)$ yields the standard Einstein--Hilbert polarization~\eqref{4.3} minus the corner term $d\vartheta^{\rm EH}$. Other notable choices are Chandrasekaran--Flanagan--Prabhu (CFP) for $(b,c)=(0,2)$~\cite{Chandrasekaran:2018aop}, Chandrasekaran--Speranza (CS) for $(b,c)=(1,2)$ \cite{Chandrasekaran:2020wwn}, and conformal, aka York, for $(b,c)=(0,1)$ \cite{Hopfmuller:2018fni,Rignon-Bret:2023fjq,Odak:2023pga}. See~\cref{boundary_conditions} for a summary.
\begin{table}[ht]
    \centering
    \begin{tabular}{c|c} 
    \hline
        $(b,c)$ & Polarization \\
        \hline
        $(0,0)$ & Corner-Free Einstein--Hilbert (EH) \\
        $(2,2)$ & Dirichlet (canonical) \\
        $(0,2)$ & Chandrasekaran--Flanagan--Prabhu (CFP) \\
        $(0,1)$ & Conformal aka York \\
        $(1,2)$ & Chandrasekaran--Speranza (CS) \\
    \hline
    \end{tabular}
    \caption{Boundary-condition choices for null-boundary polarizations. }
    \label{boundary_conditions}
\end{table}

It is important to stress that the above family of polarizations is defined for generic null hypersurfaces
and can accommodate arbitrary variations of the metric. The reason for this is that in our construction the null foliation and rigging are determined by the asymptotic fields and Bondi gauge used. Consequently, all variations are determined by the variations of the asymptotic fields. In particular, the whole set of Carrollian data, and specifically \emph{all} metric components, will vary: there is no left over gauge freedom that can be used to set $\delta k^\mu=0$, or $\delta n_\mu=0$, or $\delta \kappa=0$, for instance. Since the whole gauge is fixed, including both class I and III, the analysis of gauge dependence and anomalies of the family \eqref{thetabc} done in \cite{Odak:2023pga} does not apply. What applies instead, is the analysis of gauge dependence and  anomalies with respect to the gauge choices of $\scri$: namely, the conformal factor $r$ and the choice of foliation determined by $u$. In other words, with the bulk gauge fixing, we trade the class I and class III gauge dependence for the class II and conformal transformations. Accordingly, it is the BMS anomalies that will determine the covariant members in the $(b,c)$ family, and not the local anomalies studied in \cite{Odak:2023pga}.

\subsection{Asymptotic Limit}\label{AsymLimitPSPreSymPhysical}

Having described the two-parameter family of symplectic potentials on a finite-distance null hypersurface, we now determine how this phase-space structure behaves as the hypersurface approaches future null infinity. The null foliation constructed in~\cref{section:geometric setup} provides a controlled way to perform this limit. The purpose of this analysis is to determine whether the dependence on the polarization parameters $(b,c)$ survives in the radiative part of the potential or only through field-space exact contributions. In this way, the asymptotic limit separates the universal radiative structure from the dependence on the representative chosen at finite distance. This analysis is new and generalizes the results in~\cite{Ciambelli:2025mex}.

We recall that we are working with the Bondi condition, $\partial_u\gamma_{AB}=0$, and BMS boundary conditions
\begin{equation}\label{BMSbc}
   \delta\gamma_{AB}=0\,,\qquad \Rightarrow\qquad \delta\bF=\frac{1}{4}\delta\bR=0  \,.
\end{equation}
We also chose the leading coefficient specifying the foliation to be independent of the angular coordinates, $\partial_Ac_0=0$, and we impose $\delta c_0=0$. 
Moreover, we have $\delta\varepsilon_{\mathcal N}^{(0)}=0$, $\delta\bnabla_A=0$, and $\delta\lambda=0$.

\subsubsection{From Bulk to Boundary}\label{from bulk to boundary}

We begin by recalling the symplectic potential~\eqref{thetabc}
\begin{equation}\label{thetabc grav}
    \theta^{(b,c)} = \theta^{\rm can} +\frac{(2-b)}{16\pi G}\delta(\kappa\ve_{\cal N}) + \frac{(2-c)}{16\pi G}\delta(\theta\ve_{\cal N})\,,
\end{equation}
where
\begin{equation}
    \theta^\mathrm{can}=\varepsilon_\mathcal{N}\left(\frac{1}{2}\tau^{ab}\delta q_{ab}-\tau_a\delta\ell^a\right)\,
\end{equation}
is the $(2,2)$ Dirichlet polarization. It was shown in~\cite{Ciambelli:2025mex} that it contains the radiative part common to the entire family. We analyze it first. 

Since $\tau^{au}=0$, only the angular components contribute to the contraction $\tau^{ab}\delta q_{ab}$. The asymptotic expansions obtained in~\cref{section:geometric setup} give
\begin{align}
    \delta q_{AB} &= \lambda\delta C_{AB}+\mathcal O(\lambda^0)\,,\\
    \delta\ell^a &= \mathcal O(\lambda^{-2})\,,\\
    \tau^{AB} &= \frac{1}{16\pi G}\left(N^{AB}+2\bF\gamma^{AB}\right)\lambda^{-3}+\mathcal O(\lambda^{-4})\,,\\\label{volexp}
    \varepsilon_\mathcal{N} &= \lambda^2\varepsilon_\mathcal{N}^{(0)}+2c_0\lambda\varepsilon_\mathcal{N}^{(0)}+\mathcal O(\lambda^0)\,,
\end{align}
where
\begin{equation}
    \varepsilon_\mathcal{N}^{(0)}=\sqrt{\gamma}\,du\wedge d^2x\,.\label{e0N}
\end{equation}
Since the leading metric is fixed, varying $\gamma^{AB}C_{AB}=0$ gives $\gamma^{AB}\delta C_{AB}=0$, and we obtain
\begin{equation}
    \frac{1}{2}\tau^{ab}\delta q_{ab}\,\varepsilon_\mathcal{N}=\frac{1}{32\pi G}\varepsilon_\mathcal{N}^{(0)}N^{AB}\delta C_{AB}+\mathcal O(\lambda^{-1})\,.
\end{equation}

The second term in $\theta^\mathrm{can}$ does not contribute to the limit. Indeed, $\tau_a=\mathcal O(\lambda^{-1})$, and therefore $\tau_a\delta\ell^a\,\varepsilon_\mathcal{N}=\mathcal O(\lambda^{-1})$.
We thus find
\begin{equation}\label{theta AS}
    \lim_{\lambda\to\infty}\theta^\mathrm{can}=\frac{1}{32\pi G} \varepsilon_\mathcal{N}^{(0)}N^{AB}\delta C_{AB}=\theta^\mathrm{AS}\,, 
\end{equation}
which exactly reproduces the Ashtekar--Streubel symplectic potential at $\mathscr I^+$~\cite{Ciambelli:2025mex}. This is the polarization-independent radiative contribution to the asymptotic potential.

The novel aspect is to determine the asymptotic limit of the field-space exact terms in~\eqref{thetabc grav}. Although both $\kappa$ and $\theta$ vanish as $\lambda\to\infty$, they are multiplied by a volume form that grows as $\lambda^2$. The leading and first subleading orders must therefore be retained before taking the field-space variation. The relevant expansions are
\begin{align}
    \kappa &= -m\lambda^{-2}+\mathcal O(\lambda^{-3})\,,\\
    \theta &= -2\bF\lambda^{-1}+\left(2c_0\bF+2m+\bnabla_A\bU^A\right)\lambda^{-2}+\mathcal O(\lambda^{-3})\,.
\end{align}
Combining these expressions with the expansion of $\varepsilon_\mathcal N$ in \eqref{volexp}, we obtain
\begin{align}
    \kappa\varepsilon_\mathcal N &= -m\varepsilon_\mathcal N^{(0)}+\mathcal O(\lambda^{-1})\,,\\
    \theta\varepsilon_\mathcal N &= \left(-2\bF\lambda-2c_0\bF+2m+\bnabla_A\bU^A\right)\varepsilon_\mathcal N^{(0)}+\mathcal O(\lambda^{-1})\,.
\end{align}
The expression $\theta\varepsilon_\mathcal N$ contains a term that grows linearly with $\lambda$. However, this term depends only on quantities that are fixed on the phase space considered here. The same is true of the finite contribution proportional to $c_0\bF$. Consequently,
\begin{align}
    \lim_{\lambda\to\infty} \delta(\kappa\varepsilon_\mathcal N) = -\varepsilon_\mathcal N^{(0)}\delta m\,,\qquad \lim_{\lambda\to\infty}\delta(\theta\varepsilon_\mathcal N)=\varepsilon_\mathcal N^{(0)}\left(2\delta m-\frac{1}{2}\bnabla_A\bnabla_B\delta C^{AB}\right)\,,\label{exact limits}
\end{align}
where we used, from $\bU^A=-\frac{1}{2}\bnabla_BC^{AB}$, that
\begin{equation}\label{deltaU}
    \delta\bU^A=-\frac{1}{2}\bnabla_B\delta C^{AB}\,.
\end{equation}

Substituting these results into~\eqref{thetabc grav}, we obtain
\begin{equation}\label{physical bc limit}
    \lim_{\lambda\to\infty}\theta^{(b,c)}=\theta^\mathrm{AS}+\frac{\varepsilon_\mathcal N^{(0)}}{16\pi G}\delta\left[(b-2c+2)m+\frac{c-2}{2}\bnabla_A\bnabla_BC^{AB}\right]\,.
\end{equation}
As expected, we have shown that the polarization changes the asymptotic symplectic potential only by a field-space exact contribution and does not modify the associated presymplectic two-form. In particular, the exact contribution vanishes \emph{only} for the canonical choice $(b,c)=(2,2)$,
\begin{equation}\label{canonical limit}
    \lim_{\lambda\to\infty}\theta^{(2,2)}=\theta^\mathrm{AS}\,.
\end{equation}
We also notice that the remainders for the choices $(0,1)$ and $(0,0)$ are respectively
\be\label{theta01scri}
    \lim_{\lambda\to\infty}\theta^{(0,1)}=\theta^\mathrm{AS}+\frac{1}{16\pi G}d\delta i_\bU\varepsilon_\mathcal N^{(0)},
\ee
and
\be
    \lim_{\lambda\to\infty}\theta^{(0,0)}=\theta^\mathrm{AS}+\frac{\varepsilon_\mathcal N^{(0)}}{16\pi G}\delta\left[2m-\bnabla_A\bnabla_BC^{AB}\right]\,.
\ee
Let us compare these results with the original Wald--Zoupas analysis based on the standard Einstein--Hilbert potential.

\subsubsection{Dirichlet versus Einstein--Hilbert Potentials}

As discussed in~\cref{section:null phase spaces}, the Einstein--Hilbert representative is related to the $(b,c)=(0,0)$ member of the family by
\begin{equation}\label{EH decomposition}
    \theta^\mathrm{EH}=\theta^{(0,0)}+d\vartheta^\mathrm{EH}\,,
\end{equation}
where, after pullback to $\mathcal N$, the corner potential reads\footnote{Since $\Pi_\mu{}^a=\delta_\mu^a$, $\ell^a=n^a$ and one has $k_\mu\delta n^\mu=k_a\delta \ell^a$ on $\cN$.}
\begin{equation}
    \vartheta^\mathrm{EH}=\frac{1}{16\pi G}\left(\left(k^\mu\delta n_\mu+k_a\delta\ell^a\right)\varepsilon_\cC-k\wedge i_{\delta\ell}\varepsilon_\cC\right)\,.
\end{equation}
The two terms in this expression have different asymptotic behaviors and will therefore be evaluated separately. Notice that the wedge term has vanishing pullback to an individual cut. Nevertheless, its spacetime exterior derivative can give a finite contribution on $\mathcal N$, and the term must consequently be retained until $d\vartheta^\mathrm{EH}$ has been evaluated.

We first consider the contribution proportional to the cut area form. For the choice of normal and rigging adopted here, $k^\mu=\delta^\mu_r$ and $n_r=-1$, hence
\begin{equation}
    k^\mu\delta n_\mu=k^r\delta n_r=0\,.
\end{equation}
The remaining quantities have the asymptotic expansions
\begin{equation}
    k_a\delta\ell^a=2\delta\bbeta\,\lambda^{-2}+\mathcal O(\lambda^{-3})\,,\qquad \varepsilon_\cC=\lambda^2\varepsilon_\cC^{(0)}+\mathcal O(\lambda)\,.
\end{equation}
It follows that
\begin{equation}
    \left(k^\mu\delta n_\mu+k_a\delta\ell^a\right)\varepsilon_\cC=2\delta\bbeta\,\varepsilon_\cC^{(0)}+\mathcal O(\lambda^{-1})\,.
\end{equation}
Using $\varepsilon_\mathcal N^{(0)}=du\wedge\varepsilon_\cC^{(0)}$, its exterior derivative gives
\begin{equation}
    \lim_{\lambda\to\infty}d\left[\left(k^\mu\delta n_\mu+k_a\delta\ell^a\right)\varepsilon_\cC\right]=2\partial_u\delta\bbeta\,\varepsilon_\mathcal N^{(0)}\,.
\end{equation}
Indeed, the angular derivative of $\delta\bbeta$ wedges to zero with the top form on the two-dimensional cut, while $\varepsilon_\cC^{(0)}$ is independent of $u$. Next, from
$\bbeta=-\frac{1}{32}C^{AB}C_{AB}$
we get
\begin{equation}
    \partial_u\bbeta=-\frac{1}{16}C^{AB}N_{AB}\,.
\end{equation}
Therefore, the contribution of the first term in the corner potential is
\begin{equation}\label{first corner limit}
    \lim_{\lambda\to\infty}d\left[\left(k^\mu\delta n_\mu+k_a\delta\ell^a\right)\varepsilon_\cC\right]=-\frac{1}{8}\delta\left(C^{AB}N_{AB}\right)\varepsilon_\mathcal N^{(0)}\,.
\end{equation}

We next evaluate the wedge term. Using Cartan's identity, we find
\begin{equation}
    d\left(-k\wedge i_{\delta\ell}\varepsilon_\cC\right)=-dk\wedge i_{\delta\ell}\varepsilon_\cC+k\wedge\mathcal L_{\delta\ell}\varepsilon_\cC-k\wedge i_{\delta\ell}d\varepsilon_\cC\,.
\end{equation}
The first term is subleading because
\begin{equation}
    dk=\mathcal O(\lambda^{-2})\,,\qquad i_{\delta\ell}\varepsilon_\cC=\mathcal O(\lambda^0)\,,
\end{equation}
and hence $dk\wedge i_{\delta\ell}\varepsilon_\cC=\mathcal O(\lambda^{-2})$. The last term is also subleading. Indeed, using $d\varepsilon_\cC=\theta\ve_{\cal N}=-\theta k\wedge\varepsilon_\cC$, together with $\theta=\mathcal O(\lambda^{-1})$ and $i_{\delta\ell}k=\mathcal O(\lambda^{-2})$, we obtain
\begin{equation}
    k\wedge i_{\delta\ell}d\varepsilon_\cC=\mathcal O(\lambda^{-1})\,.
\end{equation}
The only finite contribution therefore comes from the term involving the Lie derivative. Since $\partial_Ac_0=0$, the angular components of the variation of the null generator behave as
\begin{equation}
    \delta\ell^A=\lambda^{-2}\delta\bU^A+\mathcal O(\lambda^{-3})\,.
\end{equation}
At leading order, the Lie derivative of the cut volume form is determined by the angular divergence of $\delta\ell^A$. It follows that
\begin{equation}
    k\wedge\mathcal L_{\delta\ell}\varepsilon_\cC=-\bnabla_A\delta\bU^A\,\varepsilon_\mathcal N^{(0)}+\mathcal O(\lambda^{-1})\,.
\end{equation}
Using~\eqref{deltaU},
we obtain the contribution of the wedge term,
\begin{equation}\label{second corner limit}
    \lim_{\lambda\to\infty}d\left(-k\wedge i_{\delta\ell}\varepsilon_\cC\right)=\frac{1}{2}\bnabla_A\bnabla_B\delta C^{AB}\,\varepsilon_\mathcal N^{(0)}\,.
\end{equation}

Combining~\eqref{first corner limit} and~\eqref{second corner limit}, the complete asymptotic contribution of the corner potential is
\begin{equation}\label{EH corner limit}
    \lim_{\lambda\to\infty}d\vartheta^\mathrm{EH}=\frac{\varepsilon_\mathcal N^{(0)}}{32\pi G}\left[-\frac{1}{4}\delta\left(C^{AB}N_{AB}\right)+\bnabla_A\bnabla_B\delta C^{AB}\right]\,.
\end{equation}

On the other hand, setting $(b,c)=(0,0)$ in~\eqref{physical bc limit} gives
\begin{equation}\label{00 limit}
    \lim_{\lambda\to\infty}\theta^{(0,0)}=\theta^\mathrm{AS}+\frac{\varepsilon_\mathcal N^{(0)}}{16\pi G}\left(2\delta m-\bnabla_A\bnabla_B\delta C^{AB}\right)\,.
\end{equation}
Substituting~\eqref{EH corner limit} and~\eqref{00 limit} into~\eqref{EH decomposition}, we finally obtain
\begin{align}\label{EH asymptotic limit}
    \lim_{\lambda\to\infty}\theta^\mathrm{EH}&=\theta^\mathrm{AS}-\delta b,\qquad b=-\frac{\varepsilon_\mathcal N^{(0)}}{16\pi G}\left(2m-\frac{1}{2}\bnabla_A\bnabla_BC^{AB}-\frac{1}{8}C^{AB}N_{AB}\right)\,,
\end{align}
recovering the standard result known in the literature.
It also follows from this analysis that
\begin{align}
    \delta b =\theta^{\rm AS}-\lim_{\lambda\to\infty}(\theta^{(0,0)}+ d\vartheta^{\rm EH})=\lim_{\lambda\to\infty}(\delta\ell^{(2,2)} - d\vartheta^{\rm EH}) \,.
\end{align}
As we now discuss, the identification of the boundary Lagrangian plays an important role to determine the charges.

\section{Charges}\label{sec5}

Given a symplectic potential $\th$, it is straightforward to determine the Noether current. Denoting $\hat =$ for on-shell equalities, and $\cL_{\xi}$ for the spacetime Lie derivative, we have
\be
dI_\xi\th \,{\hat =}\,\delta_\xi L =\cL_\xi L +\Delta_\xi L = d (i_\xi L+a_\xi),
\ee
where we used the fact that the only anomaly operator $\Delta_{\xi}$ acting on phase space functionals (zero forms) as $\Delta_{\xi}=\delta_\xi-\cL_\xi$,
can come from boundary terms (since the Einstein-Hilbert Lagrangian is covariant), and denoted it $a_\xi$. It follows that 
\be
j_\xi = I_\xi\theta-i_\xi L-a_\xi
\ee
is closed on-shell, and thus provides a Noether current in the presence of anomalies \cite{Freidel:2021cjp}. The second term can be omitted for tangent vector fields to the boundary, as in the BMS case, but the third term if present, can spoil the conservation and covariance properties. It is thus important to choose a covariant boundary Lagrangian, as argued in \cite{Odak:2022ndm,Rignon-Bret:2024wlu}.
Another reason comes from looking at the charges. As guaranteed by Noether's theorem,
\be\label{jxi}
j_\xi \,{\hat =}\, dq_\xi,
\ee
where $q_\xi$ is the charge aspect, and the surface integral
\begin{align}
    Q_\xi=\int_{\cal C}q_\xi
\end{align}
is the Noether charge associated with space-like hypersurfaces intersecting null infinity.
If we change the symplectic potential within the equivalence class \eqref{4.6}, say we take
\begin{align}\label{thetadecomposition}
\theta' =    \theta_{\rm EH} + \delta\ell - d\vartheta^{\rm EH}\,,
\end{align}
we find a new Noether charge with aspect
\begin{align}\label{qxiprime}
    q'_\xi = q_\xi^{\rm K} + i_\xi\ell - I_\xi\vartheta^{\rm EH}\,.
\end{align}
This formula can be used to compute the BMS charges as improved Noether charges, starting from the standard Komar expression. 
There are however two ambiguities with this procedure.  The first is that if we change the boundary Lagrangian by a corner term, this does not change the choice of symplectic potential, but affects the charge if anomalies are present \cite{Chandrasekaran:2021vyu,Odak:2022ndm}:
\begin{align}\label{psambiguity}
    \ell\to\ell + d\chi\,,\qquad \vartheta^{\rm EH}\to\vartheta^{\rm EH}+\delta\chi\,, \qquad   q^{\chi}_\xi = q'_\xi -\Delta_\xi \chi.
\end{align}
This transformation affects the chosen Noether current as well, creating a term $a_\xi=d\Delta_\xi\chi.$
The prescription of a covariant boundary Lagrangian allows one to control this  ambiguity.
The remaining ambiguity is that \eqref{jxi} only defines $q_\xi$ up to a closed 2-form, and therefore $Q_\xi$ up to a time-independent term. It can be fixed imposing covariance as in the original Ashtekar--Streubel approach, or comparing to the symplectic flow on a space-like hypersurface as in the Wald--Zoupas approach.

We now apply this procedure to determine the BMS charges starting from quasi-local charges on the null hypersurfaces $\cal N$. To determine the quasi-local charges on $\cal N$, we start from the 2-parameter family \eqref{EHtheta}, so that 
the improved Noether charge formula \eqref{qxiprime} gives
\beq\label{qimpro}
    q^{(b,c)}_\xi = q^{\rm K}_\xi + i_\xi\ell^{(b,c)} - I_\xi\vartheta^{\rm EH}\,.
\eeq
According to the discussion above, we should select the member $(2,2)$, since this gives the AS symplectic potential, and further resolve the corner ambiguity making sure that the boundary Lagrangian is covariant. In the rest of this Section we show that this procedure leads indeed to the right BMS charges.

\subsection{Phase Space Anomalies for Non-Tangent Symmetries}

In order to study the BMS charges as limits of quasi-local expressions, we need first of all a map between the BMS symmetries and diffeomorphisms at the null hypersurface $\cal N$. The Bondi framework we are using provides a natural candidate: we extend the symmetry vector fields requiring that they preserve the Bondi coordinates in the bulk. This choice gives
\begin{align}\label{BMSgeneratorExtension}
    \xi^u ={}&  f\,,\\
    \xi^r ={}& -\frac{r}{2}\bnabla_A Y^A + \frac12 \bnabla^2f+\frac{1}{r}\left(\bU^A\bnabla_A f - \frac14C^{AB}\bnabla_A\bnabla_B f\right) + {\cal O}(r^{-2})\,, \\
    \xi^A={}& Y^A - \frac{1}{r}\bnabla^A f + \frac{1}{2r^2}C^{AB}\bnabla_B f-\frac{1}{16r^3}C_{CB}C^{CB}\bnabla^A f + {\cal O}(r^{-4})\,,
\end{align}
where $f=T(x^{A})+\frac{u}{2}{\bnabla}_{A}Y^{A}$ and $Y^{A}$ is a conformal Killing vector on the sphere. With this prescription, the $\xi$'s do not preserve \emph{any} metric variations at any given $\cal N$. In fact, they are not even tangent to them:
\begin{align}\label{xin}
    (\xi^\mu n_\mu)|_{{\cal N}} = \frac12\bnabla_A Y^A \lambda +\frac12\left({c_0}\bnabla_A Y^A - \bnabla^2f -\frac12\bR f\right) + {\cal O}\left(\lambda^{-1}\right)\,.
\end{align}
They are therefore not symmetries of $\cal N$ in the sense that they do not preserve any boundary condition or universal structure there. We can nonetheless compute Noether currents and charges associated with them, since these are defined for arbitrary diffeomorphisms.\footnote{The fact that \eqref{xin} diverges at $\scri$ may give the wrong impression that $\xi$ is not tangent to $\scri$, but the well-defined normal to $\scri$ is $d\Omega=-r^{-2}dr$, whose scalar product to $\xi$ vanishes at $\scri$, see Appendix~\ref{APPCCS}.}

Consider first the Komar term in \eqref{qimpro}. The rigging vector provides a foliation of each null hypersurface $\cal N$ by 2d space-like surfaces. Any of these space-like surfaces has two null normals $(n,k)$, and the pull-back of the Komar 2-form on it can be written as follows,
\begin{align}\label{Komar}
    \underset{\longleftarrow}{q^{\rm K}_\xi} =\frac{\varepsilon_\C}{16\pi G} {\cal K}_\xi \,,\qquad
    {\cal K}_\xi = k^\mu n^\nu\nabla_\mu \xi_\nu - k_\mu n^\nu\nabla_\nu \xi^\mu = 2k^\mu n^\nu\nabla_{[\mu}\xi_{\nu]}.
\end{align}
If we decompose the diffeomorphism into tangent and non-tangent pieces to $\cal N$,
\begin{align}\label{xidec}
    \xi^\mu = \xi^\mu_\parallel - \xi_\perp k^\mu\,,\qquad \xi_\parallel^\mu n_\mu = 0\,,\qquad  \xi_\perp=\xi^\mu n_\mu,
\end{align}
we find
\begin{align}
    {\cal K}_\xi = -k^\mu \xi_\parallel^\nu W_{\mu\nu} + \alpha_k\xi_\perp + \cL_k\xi_\perp  - k_\mu n^\nu\nabla_\nu\xi_\parallel^\mu\,,
\end{align}
where
\be
\alpha_k=-n^\nu k^\mu\nabla_\mu k_\nu
\ee
is the inaffinity of the rigging.
This term is not present in the usual description of null charges \cite{Chandrasekaran:2018aop,Chandrasekaran:2021hxc,Odak:2023pga,Ciambelli:2023mir}, which refers to tangent diffeomorphisms only.
The decomposition is instead a realization of the universal corner symmetry algebra of \cite{Ciambelli:2021vnn,Ciambelli:2021nmv,Freidel:2021cjp,Ciambelli:2022cfr}. The other novelty with respect to \cite{Chandrasekaran:2018aop,Odak:2023pga} is that the non-tangential term also modifies the anomaly, and this is relevant to evaluate the third term in \eqref{qimpro}.

We now generalise the analysis of~\cite{Odak:2023pga} to see the effect of the new contribution. Consider, in the physical spacetime, a generic null hypersurface $\N$ defined by the zero-level set $\Phi=0$, and take its normal one-form to be normalized as
\begin{align}
    n_\mu = -\partial_\mu\Phi\,,
\end{align}
with rigging structure $k^\mu n_\mu = -1$ and $k^2 = 0 = n^2$\,. For instance, in our specific setting, one has $\Phi=r-r_{\cN}$, see~\eqref{normalphysical}. In general (and in particular in our setup) $\delta_\xi \Phi \neq0$\,, and its anomaly is
\begin{align}
    \Delta_\xi\Phi =\delta_\xi\Phi - \cL_\xi \Phi = \delta_\xi \Phi +\xi_\perp\,,
\end{align}
where
\begin{align}
    \xi_\perp = \xi^\mu n_\mu = -\cL_\xi\Phi\,. 
\end{align}
Then, the anomaly of the normal one-form reads
\begin{align}
    \Delta_\xi n_\mu = -\partial_\mu(\Delta_\xi\Phi)\,.
\end{align}

Using the rigging projector, we can decompose the anomaly of $n_\mu$ along its vertical and horizontal parts,
\begin{align}
    \Delta_\xi n_\mu = -w_\xi n_\mu + \zeta_\mu\,,
\end{align}
with
\begin{align}
    w_\xi := -k^\nu\partial_\nu (\Delta_\xi\Phi)\,, \qquad \zeta_\mu :=-\Pi_\mu{}^\nu\partial_\nu (\Delta_\xi\Phi)\,.
\end{align}
For the rigging vector we find
\begin{align}
    \Delta_\xi k^\mu = w_\xi k^\mu + Z_\xi^\mu\,,
\end{align}
with $Z_\xi^\mu n_\mu = 0 = Z_\xi^\mu k_\mu$\,.

Using $\delta_\xi g_{\mu\nu} =\cL_\xi g_{\mu\nu}$, we have 
\begin{align}
    k^\mu\Delta_\xi n_\mu + k_\mu\Delta_\xi n^\mu = 2w_\xi\,.
\end{align}
The covariance of $g_{\mu\nu}$ also implies that $\Delta_\xi \varepsilon_4 = 0$\,, with $\varepsilon_4=\frac1{4!}\varepsilon_{\mu\nu\rho\sigma}dx^\mu\wedge dx^\nu\wedge dx^\rho \wedge dx^\sigma$. Recalling that the cuts $\cC$ are adapted to $k$, and using $\varepsilon_\C=i_ni_k\varepsilon_4$, one has 
\begin{align}\label{iie}
    \Delta_\xi\varepsilon_\C =(i_{\zeta}i_k + i_ni_{Z_\xi})\varepsilon_4\,,
\end{align}
which is in general not zero. Nonetheless, its pullback to the cut, which is what is needed in the computation of the charges,  vanishes. To see this explicitly, consider $e_A$ and $e_B$ tangent vectors spanning the cut with $n$ and $k$ orthogonal to them. Then, the first term in~\eqref{iie} gives $(i_\zeta i_k \varepsilon_4)(e_A,e_B) = \varepsilon_4(k,\zeta,e_A,e_B)$ which vanishes as $\zeta^\mu k_\mu =0$. The second term in~\eqref{iie} also vanishes, since $(i_ni_{Z_\xi}\varepsilon_4)(e_A,e_B) = \varepsilon_4(Z_\xi,n,e_A,e_B)$, and one has $Z_\xi = Z^A_\xi e_A$\,.

The anomaly so derived enters the last term of \eqref{qimpro}, which can be written as
\begin{align}
    \underset{\longleftarrow}{I_\xi\vartheta^{\rm EH}} = {\cal A}_\xi \varepsilon_\C\,, \qquad 
    {\cal A}_\xi :={}& \frac{1}{16\pi G}(k^\mu \delta_\xi n_\mu + k_\mu\delta_\xi n^\mu)\,,
\end{align}
where we used that the pullback of $k$ on $\cC$ vanishes, and
\begin{align}\label{calA}
    {\cal A}_\xi =  \frac{1}{16\pi G}\left({\cal K}_\xi + 2 \xi^\nu k_\mu\nabla_\nu n^\mu+2w_\xi\right)\,,
\end{align}
see \eqref{Komar}. Using the decomposition \eqref{xidec}, the second term in~\eqref{calA} can be written as
\begin{align}
    \xi^\nu k_\mu \nabla_\nu n^\mu = \left(\xi^\nu_\parallel - \xi_\perp k^\nu\right)k_\mu \nabla_\nu n^\mu = \xi^a_\parallel k_b W_a{}^b - \xi_\perp \alpha_k\,.
\end{align}
We recall the Weingarten tensor $W_a{}^b=\Pi_a{}^\mu \nabla_\mu n^\nu \Pi_\nu{}^b$.
We can further split the tangential component of the diffeomorphism as
\begin{align}\label{xipar}
    \xi^\mu_\parallel = \tau_\xi n^\mu + X^\mu\,,\qquad X^\mu n_\mu = 0 =X^\mu k_\mu\,.
\end{align}
Then,
\begin{align}\label{txi}
    \xi^\mu = \tau_\xi n^\mu - \xi_\perp k^\mu + X^\mu\,,\qquad \tau_\xi=-k_\mu \xi^\mu\,,\qquad \xi_\perp = \xi^\mu n_\mu\,.
\end{align}

For the boundary Lagrangian, using $\ve_{\cal N} = -k\wedge \ve_\cC$ and
\begin{align}
    i_\xi \varepsilon_\N =\tau_\xi \varepsilon_\C + k\wedge i_\xi \varepsilon_\C\,,
\end{align}
we have
\begin{align}
    \underset{\longleftarrow}{i_\xi\ell^{(b,c)}} = -\frac{1}{16\pi G}(b\kappa+c\theta)\tau_\xi\varepsilon_\C\,.
\end{align}

Collecting all the various pieces, we arrive at the following expression for the charges evaluated at a spatial cut
\begin{align}\label{qbcCarrollGeneral}
    \underset{\longleftarrow}{q^{(b,c)}_\xi} = \left[-\xi_\parallel^ak_bT_a{}^b+\frac{\tau_\xi}{16\pi G}\left((2-b)\kappa + (2-c)\theta  \right) + \frac{1}{8\pi G}\xi_\perp\alpha_k - \frac{1}{8\pi G}w_\xi\right]\varepsilon_\C\,,
\end{align}
where we used~\eqref{NullBY}. As a consistency check, 
for tangent and anomaly free diffeomorphisms, we recover the null Brown--York charge of~\cite{Chandrasekaran:2021hxc,Ciambelli:2023mir}.
More in general,
\begin{equation}
    \underset{\longleftarrow}{q_\xi^{(2,2)}} =  \left[-\xi_\parallel^ak_bT_a{}^b+ \frac{1}{8\pi G}\xi_\perp\alpha_k - \frac{1}{8\pi G}w_\xi\right]\,\varepsilon_\cC,
\end{equation}

Returning to the general situation, we can employ~\eqref{xipar} to arrive at the expression
\begin{align}\label{qxibc}
    \underset{\longleftarrow}{q^{(b,c)}_\xi} = \frac{1}{16\pi G}\left[(2-b)\tau_\xi\kappa-c\tau_\xi\theta +2X^a\pi_a + 2\xi_\perp\alpha_k - 2w_\xi  \right]\varepsilon_\C\,,
\end{align}
which generalizes the result of~\cite{Odak:2023pga} for non-tangent vector fields. The terms proportional to $\xi_\perp$ and $w_\xi$ encode, respectively, the non-tangential component of the generator and the anomaly associated with the chosen normal. While these contributions vanish for tangent, normal-preserving symmetries, they must be retained when evaluating the charges of generic bulk extensions of BMS generators.

\subsection{Asymptotic Charges from the Null Foliation}\label{chargelimit}

The geometric decomposition of the improved charge above allows us to cleanly evaluate its asymptotic limit. First of all, given~\eqref{BMSgeneratorExtension}, the various components in~\eqref{xipar} read
\beq
    \tau_\xi &=& e^{2\beta}f=f\left(1+2\frac{\bbeta}{\lambda^2}\right)+\dots\,,\\
    X^A &=& \xi^A - fU^A - e^{2\beta}fg^{AB}\partial_B r_\lambda = Y^A - \frac{1}{\lambda}\bnabla^Af+\dots\,,\\
    \xi_\perp &=& -\xi^r + f\partial_u r_\lambda + \xi^A \partial_A r_\lambda = \lambda\left(\frac{1}{2}\bnabla_A Y^A\right) + \left(\frac12c_0\bnabla_A Y^A - \frac12 \bnabla^2 f - f\bF \right) + \dots\,,\\ 
    \alpha_k &=& 2\partial_r \beta = -\frac{4\bbeta}{\lambda^3}+\dots\,.
\eeq
For the anomaly of the normal, we have
\begin{align}
    w_\xi = -k^\mu \partial_\mu \Delta_\xi\Phi = -\partial_r\xi_\perp= -\frac12\bnabla_A Y^A - \frac{1}{\lambda^2}\left(\bU ^A\partial_A f -\frac14 C^{AB}\bnabla_A\bnabla_B f\right)+\dots\,.
\end{align}

Upon performing the asymptotic limit, $\cC$ will be a cut of $\mathscr I^+$ at fixed null time $u$. There, thanks to~\eqref{e0N}, we use the notation $\ve^{(0)}_{\cC}=i_{\pa_u}\ve^{(0)}_{\cN}=\sqrt{\gamma}d^2x$ and the expansion
\begin{align}
    \ve_\cC = \left[\l^2 + 2c_0\l + (c_0^2+2c_1)+{\cal O}(\l^{-1})\right]\ve_\cC^{(0)}~,
\end{align}
which follows directly from the Bondi determinant condition and the embedding of the null hypersurfaces. Putting things together, we get that the asymptotic value of~\eqref{qxibc} reads
\begin{align}\label{geomq}
    q_\xi^{(b,c)} =\frac{\varepsilon_\C^{(0)}}{16\pi G}&\left[\lambda^2\bnabla_A Y^A +2\lambda\left(c_0\bnabla_A Y^A+Y^A\bU _A + cf\bF \right) + (c_0^2+2c_1)\bnabla_A Y^A \right. \nonumber \\&+2c_0Y^A\bU _A +2cc_0f\bF  + (b-2c-2)fm + \frac{c}{2}f\bnabla_A\bnabla_BC^{AB}-\frac12C^{AB}\bnabla_A\bnabla_B f \nonumber \\&\left.- 2Y^A\left(\bP _A + 2\bnabla_A\bbeta -\bnabla_A c_1\right)-4\bnabla_A Y^A\bbeta \right] + {\cal O}(\lambda^{-1})\,.
\end{align}

This result can be cross-checked starting directly from~\eqref{qimpro} and evaluating the asymptotic limit of each term in the right-hand side. While this obscures the geometric decomposition, it provides a useful independent verification. The first term in~\eqref{qimpro} is the Komar two-form, and has the asymptotic limit
\begin{align}
    q_\xi^{\rm K}={}& \frac{\varepsilon_{\cal C}^{(0)}}{16\pi G}\Bigg[\lambda^2 \bnabla_A Y^A + 2\lambda(c_0\bnabla_A Y^A + Y^A\bU _A) + (c_0^2+2c_1)\bnabla_A Y^A + 2c_0 Y^A\bU _A \nonumber \\ {}&- f(2m+\frac18C_{AB}N^{AB})-\frac14 C^{AB} \bnabla_A\bnabla_B f - 2Y^A(\bP_A + \partial_A\bbeta - \partial_A c_1) -2\bnabla_A Y^A\bbeta\Bigg]+{\cal O}(\lambda^{-1})\,.
\end{align}
For the boundary Lagrangian contribution, we find
\begin{align}\label{ixilbc}
i_\xi\ell^{(b,c)}= \frac{\varepsilon_{\cal C}^{(0)}}{16\pi G}f\left[2c\bF \lambda + (b-2c) m + 2cc_0\bF + \frac{c}{2} \bnabla_A\bnabla_B C^{AB}\right]+{\cal O}(\lambda^{-1})\,.
\end{align}
Eventually, the corner term takes the form
\begin{align}
    I_\xi \vartheta^{\rm EH} = \frac{\varepsilon_{\cal C}^{(0)}}{16\pi G}\left[2f\dot{\bbeta} + 2Y^A\bnabla_A\bbeta + 2\bnabla_A Y^A \bbeta + \frac14C^{AB}\bnabla_A\bnabla_B f\right]+{\cal O}(\lambda^{-1})\,.
\end{align}
Adding up,~\eqref{qimpro} reproduces exactly~\eqref{geomq}.

After integration on the cuts, several terms integrate to zero,\footnote{Here, we use  $\bnabla_{(A}Y_{B)} = \frac12 \bnabla_C Y^C \gamma _{AB}$ and the tracelessness of the shear to render $\lambda Y^A \bU_A$ a total derivative.} and we find
\begin{align}\label{improvnoetheronshell}
    Q^{(b,c)}_\xi=-\int_{\cC} \frac{\varepsilon_{\cal C}^{(0)}}{16\pi G}\left[2cf\left(\lambda+c_0\right)\bF  + f\left((b-2c-2)m+\frac{c-1}{2}\bnabla_A\bnabla_B C^{AB}\right) - 2Y^A\bP _A\right]+{\cal O}(\lambda^{-1})\,,
\end{align}
where the global minus sign comes from the orientation of the cut, viewed as the outer boundary of a spacelike hypersurface with future-directed normal. In other words, the cut is oriented as the outward boundary of a spacelike hypersurface, which in our conventions is opposite to the orientation induced by $\ve_\cC=i_\ell i_k \ve_4$.

In particular, for the Einstein--Hilbert representative $b=0=c$, one has
\begin{align}
    Q^{(0,0)}_\xi ={}&\int_{\cC} \left(q^{\rm K}_\xi - I_\xi \vartheta^{\rm EH}\right) =  \int_{\cC}\frac{\varepsilon^{(0)}_{\cal C}}{16\pi G}\left[2fm+\frac{1}{2}f\bnabla_A\bnabla_B C^{AB} +2Y^A\bP _A\right]+{\cal O}(\lambda^{-1})\,,
\end{align}
which matches~\cite{Odak:2022ndm}. Notice that the last term in \eqref{ixilbc} is a corner contribution, see e.g. \eqref{theta01scri}, hence studying the different members in this family reproduces the analysis of \cite{Chandrasekaran:2021vyu}.

The first term in \cref{improvnoetheronshell} diverges. However, one remarks that it only depends on the chosen asymptotic conformal frame and on the location of the regulator. Therefore, it is already present in Minkowski spacetime. Indeed, for $m=\bbeta=\bU ^A=\bP ^A = C_{AB} = 0$, the boundary Lagrangian reads 
\begin{align}
    \ell^{(b,c)}_{\rm Mink} = \frac{\ve^{(0)}_{\cN}}{16\pi G} 2c(\lambda+c_0)\bF + O(\lambda^{-1})\,.
\end{align}
It thus represents a vacuum reference contribution that can be removed before taking the asymptotic limit.

We therefore define the renormalized boundary Lagrangian as
\begin{align}\label{renbcbdrylag}
    \ell^{(b,c)}_{\rm Ren} = \ell^{(b,c)}- \ell^{(b,c)}_{\rm Mink}\,.
\end{align}
Note that this does not impact the symplectic potential, since $\delta \ell^{(b,c)}_{\rm Mink}$ vanishes asymptotically. The renormalized charge is then given by
\begin{align}
    q^{(b,c)}_{\xi,\rm Ren} :=  q^{\rm K}_\xi + i_\xi\ell^{(b,c)}_{\rm Ren} - I_\xi \vartheta^{\rm EH}\,.
\end{align}
Then, we obtain
\begin{align}\label{Qren}
    Q^{(b,c)}_{\xi,\rm Ren}= -\frac{1}{16\pi G}\int_{\cal C}\left[(b-2c-2)fm+\frac{c-1}{2}f\bnabla_A\bnabla_B C^{AB}-2Y^A\bP _A\right]\varepsilon_{\cal C}^{(0)}+{\cal O}(\lambda^{-1})\,.
\end{align}
Note that, at the level of integrated charges, this prescription is equivalent to subtracting the complete charge of Minkowski, since the other contributions are total derivatives. Therefore
\begin{align}
    Q^{(b,c)}_{\xi,\text{Ren}}:= Q^{(b,c)}_{\xi} - Q^{(b,c)}_{\xi, \rm Mink}\,.
\end{align}
This type of subtraction is familiar from other studies of limits of quasi-local quantities, see e.g. \cite{Brown:1992br,Ashtekar:2024xwl,Szabados:2009eka}. 

We now select the Dirichlet representative, since as previously established this is the one that corresponds to the AS symplectic potential. This gives
\begin{align}\label{DirichletRenCharge}
    Q^{(2,2)}_{\xi,\text{Ren}}= \frac{1}{16\pi G}\int_{\cal C}\left[4fm + 2Y^A\bP _A -\frac12 f \bnabla_A\bnabla_BC^{AB}\right]\varepsilon_{\cal C}^{(0)},
\end{align}
which differs from the correct BMS charges \cite{Ashtekar:1981bq,Dray:1984rfa,Wald:1999wa,Barnich:2010eb,Flanagan:2015pxa,Rignon-Bret:2024gcx},\footnote{The angular momentum aspect $N^A$ of~\cite{Barnich:2010eb} is related to our $\bP ^A$ by $N^A = \bP ^A +  \bnabla^A\bbeta = \bP ^A - \frac{1}{32}\bnabla^A\left(C_{BC}C^{BC}\right)$.} by the extra soft term $-\frac12 f \bnabla_A\bnabla_BC^{AB}$. 
Thus, matching symplectic potential together with finiteness and background subtraction are not sufficient to select the right charges. 
Following~\cite{Odak:2022ndm}, we will show in the next subsection that the remaining difference originates from the non-covariance of the Dirichlet boundary Lagrangian. Restoring covariance determines an anomalous change of corner representative, which precisely removes the soft term and yields the standard BMS charges.

\subsection{Corner Improvement of BMS Charges}\label{cornBMS}

We now determine the corner improvement needed to recover the
standard BMS charges from the renormalized Dirichlet ones.
The Wald--Zoupas prescription selects a preferred boundary
symplectic potential by requiring, in particular, covariance
and vanishing flux in stationary configurations
\cite{Wald:1999wa}. In our setting, the canonical $(2,2)$
polarization already yields this preferred radiative
potential, namely the Ashtekar--Streubel one. However, the
corresponding improved Noether charges need not coincide
with the Wald--Zoupas charges, since the boundary Lagrangian
$\ell$ may carry a soft anomaly, $\Delta_\xi\ell\neq 0$,
which contributes to the flux-balance law. As shown in
\cite{Odak:2022ndm}, for Wald--Zoupas-compatible soft
anomalies, the preferred charges can be recovered as
improved Noether charges by selecting a corner representative (as discussed in equation \eqref{psambiguity})
that renders the boundary Lagrangian covariant.
In our case, this amounts to finding a corner improvement
$\chi$ such that the shifted boundary Lagrangian
$\ell_{\rm Cov}=\ell_{\rm Ren}+d\chi$ satisfies
$\Delta_\xi\ell_{\rm Cov}\doteq 0$, where equality is
understood up to total derivatives on the cuts.
This shift leaves the symplectic potential unchanged,
while modifying the Noether charges by precisely the
soft contribution. We now construct this improvement
and show that it leads to the standard BMS charges.

Using~\eqref{renbcbdrylag} in the canonical polarization, one finds
\begin{align}
    \ell^{(2,2)}_{\rm Ren}= \frac{\varepsilon^{(0)}_{\cN}}{16\pi G}\left(-2m+\bnabla_A\bnabla_B C^{AB}\right)+{\cal O}(\lambda^{-1})\,.
\end{align}

To study its covariance, we evaluate $\Delta_\xi \ell^{(2,2)}_{\rm Ren}$. At the asymptotic boundary, we find\footnote{All expressions from now on are evaluated at the asymptotic boundary, $\lambda\to \infty$.}
\begin{align}
    \Delta_\xi\ell^{(2,2)}_{\rm Ren}=-\frac{1}{8\pi G}\left[\Delta_\xi(m\varepsilon^{(0)}_\cN) + \Delta_\xi\left(\bnabla_A\bU ^A\varepsilon^{(0)}_\cN\right)\right]\,.
\end{align}

In order to compute this, we use the anomalies of Bondi data and the derivation given in Appendix B of~\cite{Odak:2022ndm}. Using
\begin{align}
    \Delta_\xi \varepsilon^{(0)}_\cN ={}&-3\partial_u f\varepsilon^{(0)}_\cN\,,\\
    \Delta_\xi m ={}& 3\partial_u f m + \frac12 \bnabla_A N^{AB}\bnabla_B f + \frac14 \partial_u (C^{AB}\bnabla_A\bnabla_B f)\,,\\
    \Delta_\xi C_{AB} ={}& -\pa_u f C_{AB}-2\bnabla_{\langle A}\partial_{B\rangle}f\,,
\end{align}
we have
\begin{align}
    \Delta_\xi(m\varepsilon^{(0)}_\cN) = \left[\frac12\bnabla_A N^{AB}\bnabla_B f + \frac14 \partial_u (C^{AB}\bnabla_A\bnabla_B f)\right]\varepsilon^{(0)}_\cN\,,
\end{align}
and
\begin{align}
    \Delta_\xi\left(\bnabla_A \bU ^A\varepsilon^{(0)}_\cN \right) = \left[\bnabla_A\left(\left(\Delta_\xi - 3\partial_u f\right)\bU ^A\right) + \partial_u \bU ^A\bnabla_A f + \bU ^A\bnabla_A\partial_u f\right]\varepsilon^{(0)}_\cN\,.
\end{align}
Since we are eventually interested in integrated expressions over the cut, we introduce the notation $\doteq$ to indicate equalities up to total derivatives on the cut. Then, we obtain the compact expression
\begin{align}
    \Delta_\xi\left(\bnabla_A\bU ^A\varepsilon^{(0)}_\cN\right) \doteq \frac12 \partial_u(C^{AB}\bnabla_A\bnabla_B f)\varepsilon^{(0)}_\cN\,,
\end{align}
such that the total anomaly reduces to
\begin{align}
    \Delta_\xi\ell^{(2,2)}_{\rm Ren} \doteq -\frac{\varepsilon^{(0)}_\cN}{16\pi G}\left[\bnabla_A N^{AB}\bnabla_B f + \frac32\partial_u (C^{AB}\bnabla_A\bnabla_B f)\right]\,.
\end{align}

Now, we restrict our attention to the BMS group on the round sphere, for which $\bnabla_{\langle A}\bnabla_{B\rangle}\partial_u f = 0$. This then implies\footnote{For a global conformal Killing vector on the round sphere, $\bnabla_{(A}Y_{B)}=\frac12\bnabla_C Y^C\gamma_{AB}$ implies $\bnabla_{\langle A}\bnabla_{B\rangle}\bnabla_C Y^C=0$. Since $\partial_u f=\bnabla_C Y^C/2$ and $C^{AB}$ is traceless, $C^{AB}\bnabla_A\bnabla_B\partial_u f=\frac12 C^{AB}\bnabla_{\langle A}\bnabla_{B\rangle}\bnabla_C Y^C=0$.
} $C^{AB}\bnabla_A\bnabla_B \partial_u f = 0$ and thus
\begin{align}
    \partial_u (C^{AB}\bnabla_A\bnabla_B f) = N^{AB}\bnabla_A\bnabla_B f\,.
\end{align}
Then, we find
\begin{align}
    \Delta_\xi \ell^{(2,2)}_{\rm Ren} \doteq -\frac{\varepsilon^{(0)}_\cN}{32\pi G}\partial_u \left(C^{AB}\bnabla_A\bnabla_Bf\right)\,.
\end{align}

Now, if we introduce the two-form
\begin{align}
    s_\xi:=\frac{\varepsilon_{\cal C}^{(0)}}{32\pi G}C^{AB}\bnabla_A\bnabla_B f \,,
\end{align}
we have
\begin{align}
    d s_\xi\doteq -\Delta_\xi \ell^{(2,2)}_{\rm Ren} \,.
\end{align}
Then, using the corner improvement discussed in~\eqref{psambiguity},
\begin{align}
    \ell^{(2,2)}_{\rm Cov} = \ell^{(2,2)}_{\rm Ren} + d\chi\,,\qquad \vartheta^{\rm EH}_{\rm Cov} = \vartheta^{\rm EH} + \delta\chi\,,
\end{align}
the anomaly of the shifted boundary Lagrangian is a priori
\begin{align}
    \Delta_\xi\ell^{(2,2)}_{\rm Cov} = \Delta_\xi\ell^{(2,2)}_{\rm Ren} + d\Delta_\xi\chi \doteq - d s_\xi + d\Delta_\xi\chi\,.
\end{align}
Therefore, a covariant representative is obtained by choosing
\begin{align}
    \Delta_\xi \chi \doteq s_\xi\,.
\end{align}

To identify a local solution, we observe that
\begin{align}
    s_\xi\doteq \frac{1}{16\pi G}\bU ^A\bnabla_A f \varepsilon_{\cal C}^{(0)}\,.
\end{align}
This suggests using the corner improvement
\begin{align}
    \chi= -\frac{1}{16\pi G}d u\wedge i_{\bU }\varepsilon_{\cal C}^{(0)}\,.
\end{align}
Indeed, using $\Delta_\xi d u = - d f$ and evaluating on the cut, one has
\begin{align}
    \underset{\longleftarrow}{\Delta_\xi \chi} = \frac{1}{16\pi G}df \wedge i_{\bU }\varepsilon_{\cal C}^{(0)} = \frac{1}{16\pi G}\bU ^A\bnabla_A f \varepsilon_{\cal C}^{(0)} \doteq s_{\xi}\,.
\end{align}
Notice that this establishes the desired equality only after pullback to a cut. More generally, the difference
$\Delta_\xi\chi-s_\xi$ may contain terms proportional to $du$, as well as cut-exact contributions. Upon taking
an exterior derivative, these terms contribute only total derivatives on the cuts. They therefore do not affect the covariance condition in the integrated sense relevant for the charges and fluxes considered here.
We thus have obtained
\begin{align}
    \Delta_\xi\ell^{(2,2)}_{\rm Cov} = \Delta_\xi\left(\ell^{(2,2)}_{\rm Ren}+d\chi \right) \doteq 0\,.
\end{align}

As explained in \eqref{psambiguity}, this corner improvement modifies the Noether charge as in~\eqref{psambiguity}. Therefore, the final integrated covariant charge reads
\begin{align}
    Q^{(2,2)}_{\xi,\chi}= Q_{\xi,\rm Ren}^{(2,2)}- \int_{\cal C}\Delta_\xi \chi= \frac{1}{16\pi G}\int_{\cal C}\left(4fm +2Y^A\bP _A  \right)\varepsilon_{\cal C}^{(0)}\,.
\end{align}
This now exactly matches the BMS charges found in~\cite{Barnich:2010eb}, which is the main result of this section. This shows that the direct asymptotic limit of the physical $(2,2)$ charge and the standard BMS charge belong to different corner representatives. Their difference is removed by the local corner two-form $\chi$, which is determined independently by requiring covariance of the limiting boundary Lagrangian. The cancellation of the soft term is not an ad hoc modification of the charge, but a consequence of selecting the covariant representative prescribed by the Wald--Zoupas construction. The representative remains unique only up to the addition of anomaly-free corners and cut-exact forms, neither of which changes the integrated charge.

We finally consider the asymptotic flux associated with the canonical polarization. As shown in \cref{from bulk to boundary}, its symplectic potential reduces directly to the Ashtekar--Streubel potential. Moreover, the Minkowski reference term has vanishing field-space variation within the asymptotic phase space considered here, and therefore does not modify this limit. We thus
obtain
\begin{align}
    \lim_{\lambda\to\infty}I_\xi \theta^{(2,2)}_{\rm Ren} = I_\xi \theta_{\rm AS} = \frac{\varepsilon_{\cal N}^{(0)}}{32\pi G}N^{AB}\delta_\xi C_{AB}\,.
\end{align}
Thus, the finite distance flux reduces directly to the Ashtekar--Streubel flux at null infinity~\cite{Ashtekar:1981bq}. Moreover, using $\Delta_\xi \ell^{(2,2)}_{\rm Ren} \doteq -ds_\xi$ together with the corner improvement $\Delta_\xi \chi \doteq s_\xi$, the asymptotic Noether identity becomes
\begin{align}
    \lim_{\lambda\to\infty}I_\xi \theta^{(2,2)}_{\rm Ren} \doteq d\left(q^{(2,2)}_{\xi,\rm Ren} - s_\xi\right) \doteq dq_{\xi,\chi}^{(2,2)}\,.
\end{align}
Hence, the same corner representative that renders the limiting boundary Lagrangian covariant also reproduces the standard BMS flux-balance law. This result complements the gravitational dynamics studied in Section \ref{sec3}: the evolution of the Bondi mass and angular-momentum aspects entering the charges is governed by the asymptotic Raychaudhuri and Damour constraints, respectively. The asymptotic evolution equations, charges and radiative flux thus emerge consistently from the same finite-distance null construction.

In summary, we have shown that the finite-distance $(2,2)$ improved Noether charges admit a well-defined asymptotic limit after subtracting a background contribution. The resulting boundary Lagrangian is rendered covariant by a corner representative fixed up to anomaly-free contributions, such that the residual soft term is removed and one exactly reproduces the standard Bondi charges.

\subsection{Conformal polarization}

Looking at the coefficient of the mass term in \eqref{Qren}, we recognize the usual ``factor of 2'' problem with the Komar mass for $b=c=0$. We also notice that insofar as this problem is concerned, we have two options to fix it: $(2,2)$ and $(0,1)$, given respectively by the Dirichlet polarization
\eqref{DirichletRenCharge}, and the conformal polarization
\begin{align}\label{Q01}
    Q^{(0,1)}_{\xi,\text{Ren}}= \frac{1}{16\pi G}\int_{\cal C}\left[4fm + 2Y^A\bP _A \right]\varepsilon_{\cal C}^{(0)}.
\end{align}
This second option reproduces directly the correct BMS charges. The reason why it is not selected by the generalized Wald-Zoupas prescription is that \eqref{theta01scri} fails to satisfy both covariance and stationarity. On the other hand, $I_\xi\theta^{(0,1)}$, does satisfy covariance and stationarity, has a covariant boundary Lagrangian, and that explains why \eqref{Q01} gives the right charge:
\be
j_\xi^{(0,1)} = I_\xi \theta^{(0,1)} \doteq I_\xi \th^{AS} \,{\hat =}\, dq^{BMS}_\xi
\ee
where in the second equality we used $\delta_\xi(\bnabla \bU\epsilon_\scri)\, \dot{=}\, 0$. It is easy to check that it is the only member of the 2-parameter family with these properties.
In other words, if one weakens the prescription of \cite{Odak:2022ndm} and requires stationarity and covariance for $\ell$ and $I_\xi\th$, as opposed to for the full $\theta$, one finds \emph{two} possible choices of symplectic potentials: $\th^{AS}=\th^{(2,2)}$, and $\th^{(0,1)}$. The two give the same symplectic structure, and same charges and fluxes for the BMS group.

\section{Final Words}\label{conclusions}

In this work, we have investigated how the geometry, dynamics, gravitational phase space, and charges of finite-distance null hypersurfaces approach future null infinity. Starting from a family of null hypersurfaces constructed in Bondi gauge, we used the rigging formalism to study the asymptotic limit of their intrinsic Carrollian geometry. Our analysis was carried out primarily in the physical spacetime, complemented by a parallel derivation of the gravitational constraints and symplectic potentials in the conformally compactified spacetime, presented in Appendix~\ref{APPCCS}. This provides a common framework for relating gravitational dynamics and observables at finite distance to their familiar asymptotic counterparts.

We first studied the gravitational constraints intrinsic to null hypersurfaces. Extending the previous analysis of~\cite{Ciambelli:2025mex}, we showed that not only the Raychaudhuri equation reproduces the Bondi mass-loss formula, but also that the Damour equation yields the angular-momentum evolution equation in the asymptotic limit. This correspondence holds in both the physical and conformally compactified descriptions, although the individual geometric contributions are organized differently. In particular, the conformally compactified treatment shows how the effective Ricci contributions generated by the conformal transformation participate in recovering the Bondi evolution equations. From this perspective, both equations governing the asymptotic gravitational dynamics arise as limits of the ordinary gravitational constraints intrinsic to finite-distance null hypersurfaces.

We then turned to the gravitational phase space. Starting from the two-parameter family of null-boundary symplectic potentials labelled by $(b,c)$, we determined their asymptotic behaviour and identified the role of the different boundary polarizations. Within the phase space considered here, the canonical $(2,2)$ polarization reduces to the Ashtekar--Streubel potential, while the remaining members of the family differ from it only by field-space exact contributions. Consequently, all these representatives define the same asymptotic radiative presymplectic structure, despite retaining information about the choice of boundary Lagrangian. We further showed that the conformally compactified description reproduces exactly the same asymptotic family, although the conformal transformation redistributes the contributions from the Einstein--Hilbert potential, boundary Lagrangian, and corner term. This comparison makes explicit how different finite-distance polarizations and geometric descriptions lead to the same radiative presymplectic structure at null infinity.

Finally, we investigated the corresponding gravitational charges. An important subtlety is that generic bulk extensions of BMS generators are not tangent to the finite-distance null hypersurfaces, requiring us to account for their normal components and the associated phase-space anomalies. We derived a geometric expression for the improved Noether charges and evaluated its asymptotic limit along the physical null foliation. After subtracting a divergent Minkowski reference contribution, we obtained a finite family of charges depending on the polarization parameters. The canonical $(2,2)$ polarization yields the correct normalization of the Bondi charges, but retains an additional soft contribution. Following the Wald--Zoupas prescription and its formulation in terms of improved Noether charges, we showed that requiring covariance of the limiting boundary Lagrangian, up to cut-total derivatives, determines a local corner improvement that removes this contribution. The resulting charges reproduce the standard BMS expressions, while the radiative symplectic potential remains unchanged and the corresponding flux-balance law is recovered. Thus, the asymptotic limit of the symplectic structure alone does not completely determine the charges: the choice of boundary and corner representatives plays an essential role in identifying their covariant asymptotic expressions.

Taken together, these results establish a direct correspondence between finite-distance null physics and the gravitational structures at null infinity, encompassing the intrinsic Einstein constraints, radiative phase space, and BMS charges. They also show that taking the asymptotic limit involves more than retaining the leading terms of an expansion: boundary Lagrangians, corner contributions, and phase-space anomalies can leave finite imprints that are essential for identifying the appropriate asymptotic observables. In particular, while the radiative presymplectic structure is universal within the phase space considered here, the identification of the corresponding gravitational charges requires a careful choice of boundary and corner representatives. This distinction emerges naturally from our finite-distance construction and highlights the importance of these ambiguities in connecting bulk and asymptotic gravitational physics.

Several directions remain open. A natural next step is to relax the boundary conditions imposed here, particularly the restriction to a fixed leading metric on the cuts, and explore generalized and extended BMS symmetries within the same geometric framework. Another direction concerns the renormalization of gravitational observables at null infinity. Our construction provides a finite-distance interpretation of the subtractions and corner improvements entering the asymptotic charges, which may offer useful insights into holographic renormalization in asymptotically flat gravity. In this respect, the curvature-dependent trace of the rescaled null Brown--York stress tensor, discussed in \cref{NBY}, suggests a possible connection with trace-anomaly-like structures in celestial holography. It would also be interesting to investigate subleading corrections to the asymptotic charges and their relation to gravitational memory and soft theorems, thereby extending the correspondence beyond the leading asymptotic structures studied here. Finally, a more far-reaching direction is to explore whether finite-distance null geometry and its associated charges can provide a geometric starting point for gravitational light-ray operators and conformal-collider observables, connecting finite-distance gravitational physics with celestial energy correlators.

\paragraph{Acknowledgments} We are thankful to Gabriel Arenas-Henriquez, Diego Fernández-Silvestre, Laurent Freidel, Hernán González, Temple He, Sercan Hüsnügil, Marc Klinger, Puttarak Jai-akson, Fabio Mele, Giulio Neri, A.W. Peet, Javi Peraza, David Rivera-Betancour, Per Sundell, Tom Wetzstein, and Kathryn Zurek for useful discussions. The work of F.D. was supported by FONDECYT Grant No. 3260880. L.S. acknowledges the partial support from Fondecyt Regular No. 1240043 and by the Beca Doctorado Nacional ANID Grant No. 21221813. M.I. is also supported by the Discovery Grant from the Natural Sciences and Engineering Research Council of Canada. M.A. acknowledges support from grant PID2023-148373NB-I00, funded by MCIN/AEI/10.13039/501100011033/FEDER--UE. M.A., M.I., L.S., and S.S. thank Perimeter for hospitality during various stages of this project. L.S. thanks Tsinghua University and in particular Wei Song for the invitation to present this work. Research at Perimeter Institute is supported in part by the Government of Canada through the  Department of Innovation, Science and Economic Development Canada and by the Province of  Ontario through the Ministry of Colleges and Universities.
\begin{appendix}

\section{Conformally Compactified Spacetime}\label{APPCCS}

In this appendix, we present a complementary analysis of the geometry, intrinsic gravitational constraints and symplectic potentials in the conformally compactified spacetime. This parallels the physical-spacetime analysis of \cref{section:geometric setup,Sec:GravPS}, providing a different derivation of the same asymptotic results. In this description, $\mathscr I^+$ is an actual null hypersurface, and the conformal transformation reorganizes the intermediate geometric and symplectic contributions.

To carry out this analysis, we bring $\mathscr I^+$ to finite coordinate distance by conformally compactifying the spacetime $({\cal M},g)$.\footnote{The conformal boundary of asymptotically flat spacetimes is a disjoint pair of two null hypersurfaces $\mathscr{I}^{\pm}$, but we only focus on $\mathscr{I}^{+}$.} This makes the notion of asymptotic infinity, now viewed as an embedded hypersurface in a larger auxiliary manifold, well-defined, and given by the hypersurface located at the set of points where $\Omega=0$. In the physical spacetime one expands metric components in powers of $1/r$, capturing radiative degrees of freedom at subleading orders. In the conformal spacetime, this corresponds to expanding in powers of $\Omega$, treated as a coordinate on ${\cal M}$. We first write the conformally compactified metric, expand it in powers of $\Omega$ and perform an asymptotic expansion of the intrinsic Carrollian data analogous to that of \cref{section:geometric setup}.

\subsection{Geometric Setup in Bondi Gauge}

Let us make the following coordinate transformation
\begin{equation} \label{2.48}
    \Omega=\frac1{r} \,, \qquad d r=-\frac{d \Omega}{\Omega^2}\,.
\end{equation}
The line element \eqref{2.1} then becomes
\begin{equation} \label{2.49}
    ds^{2} = -2e^{2\beta}du\left(-\frac{d\Omega}{\Omega^2}+Fdu\right) + g_{AB}(dx^{A}-U^{A}du)(dx^{B}-U^{B}du)\,,
\end{equation}
with 
\begin{align} 
    F &= \bF  - m \Omega + F_2 \Omega^2+\mathcal{O}(\Omega^{3})\,,  \label{2.50}\\
    \beta &= \bbeta\Omega^2+\mathcal{O}(\Omega^{4}) \,,\label{2.51}  \\
    U^{A} &= \bU^{A}\Omega^2 + U^{A}_{1}\Omega^3 +\mathcal{O}(\Omega^{4})\,, \label{2.52} \\
    g_{AB} &= \frac{\gamma_{AB}}{\Omega^2} + \frac{C_{AB}}{\Omega} + \frac{1}{4}\gamma_{AB}C_{CD}C^{CD} + \mathcal{O}(\Omega^{1})\,. \label{2.53}
\end{align} 
In the physical spacetime, the boundary is reached in the limit $r\to\infty$. Under the coordinate transformation~\eqref{2.48}, this translates into the limit $\Omega\to0$, thus putting the conformal boundary at the location $\Omega=0$. To remove the second-order pole of the line element~\eqref{2.49} at $\Omega=0$, we rescale it by a factor of $\Omega^{2}$, obtaining the unphysical (or conformal) metric $\hat{g}_{\mu\nu} = \Omega^{2}g_{\mu\nu}$. In terms of the new coordinates $x^{\mu} = (\Omega,x^{a})=(\Omega,u,x^{A})$, we have
\beq
    d\hat{s}^{2}=\Omega^{2}ds^{2}  =2e^{2\beta}du\left(d\Omega-\Omega^2 Fdu\right) + \Omega^2  g_{AB}(dx^{A}-U^{A}du)(dx^{B}-U^{B}du)\,.\label{unphys}
\eeq

In the conformally-compactified geometry~\eqref{unphys}, $\mathscr{I}^{+}$ is the embedded null hypersurface corresponding to the set of points such that $\Omega=0$ while $d\Omega\neq0$. Evaluating the ambient conformal metric at $\Omega=0$ gives
\begin{equation} \label{2.55}
    d\hat{s}^{2}\vert_{\Omega=0}= 0\,du^{2}+2d u\, d\Omega+\gamma_{AB}dx^{A}dx^{B}\,,
\end{equation}
such that the induced metric on $\mathscr{I}^{+}$ is
\begin{equation}
    \underset{\longleftarrow}{d\hat{s}^{2}}=0\,du^{2}+\gamma_{AB}dx^{A}dx^{B}\,.
\end{equation}
From this we see both the degeneracy of the induced metric on ${\mathscr{I}}^{+}$ and that $\gamma_{AB}$ is indeed the metric on a $u=\mathrm{const}$ cut of $\mathscr{I}^{+}$. From~\eqref{unphys}, we can expand the components of the metric in powers of $\Omega$, according to the boundary conditions in \cref{2.50,2.53}. This yields the following expansions
\begin{align}
    \hat{g}_{uu}&=-2\bF\Omega^2+2m\Omega^3+\Omega^4(\bU^2-4\bbeta\,\bF-2F_2)+\cO(\Omega^5)\,, \label{2.56} \\
     \hat{g}_{u A} &= -
    \Omega^2 \left( \bU _A+\Omega (U_{1A}+C_{AB}\bU ^{B})\right)+\cO(\Omega^4)\,, \label{2.57} \\
    \hat{g}_{u\Omega} &= 1+2\bbeta \Omega^2+\cO(\Omega^3)\,, \label{2.58}  \\
    \hat{g}_{AB} &=\gamma_{AB}+\Omega C_{AB} +\frac14 \Omega^2\gamma_{AB}C_{CD}C^{CD}+\cO(\Omega^3) \,,\label{2.59}
\end{align}
with all other metric components being zero. One can perform a similar expansion for the inverse metric, 
\begin{align}
    \hat{g}^{\Omega\Omega} &= 2\Omega^{2}\bF  - 2\Omega^{3}m + {\cal O}(\Omega^{4})\,, \label{2.60}\\
    \hat{g}^{\Omega u} &= 1-2\bbeta\Omega^2 + {\cal O}(\Omega^4)\,, \label{2.61} \\ 
    \hat{g}^{\Omega A} &= \bU^A\Omega^2 + U_1^A\Omega^3 + {\cal O}(\Omega^4)\,. \label{2.62}
\end{align} 

We conclude this overview of the conformally-compactified line element stressing that the Einstein equations are not conformal. Indeed, the Ricci tensor transforms non-linearly under a conformal rescaling of the metric $\hat g_{\mu\nu}=\Omega^2 g_{\mu\nu}$ as 
\begin{equation} \label{2.63}
    R_{\mu\nu} = \hat{R}_{\mu\nu} + \frac{2}{\Omega}\hat{\nabla}_{\mu}\pa_{\nu}\Omega + \hat{g}_{\mu\nu}\hat{g}^{\alpha\beta}\left(\Omega^{-1}\hat{\nabla}_{\alpha}\pa_{\beta}\Omega - 3\Omega^{-2}\pa_{\alpha}\Omega\pa_{\beta}\Omega\right)\,,
\end{equation}
where $\hat{\nabla}$ is the bulk Levi-Civita connection associated to the conformal metric $\hat{g}_{\mu\nu}$. Since we assume that the physical Bondi--Sachs metric satisfies the vacuum Einstein equations, $R_{\mu\nu}=0$, the Ricci tensor for the conformally compactified metric is
\begin{equation} \label{hatRicciTensor}
    \hat{R}_{\mu\nu} = -\frac{2}{\Omega}\hat{\nabla}_{\mu}\pa_{\nu}\Omega - \hat{g}_{\mu\nu}\hat{g}^{\alpha\beta}\left(\Omega^{-1}\hat{\nabla}_{\alpha}\pa_{\beta}\Omega - 3\Omega^{-2}\pa_{\alpha}\Omega\pa_{\beta}\Omega\right)\,,
\end{equation}
which displays the effective Ricci contributions generated by the conformal factor. Our next goal is to describe how to foliate the conformal Bondi spacetime~\eqref{unphys} by a family of null hypersurfaces, just as we did in the physical spacetime.

\subsubsection*{Asymptotic Limit of Null Foliation}

Let $\cal N$ be a general embedded hypersurface in our conformal spacetime (${\cal M},\hat{g}$). As in the physical spacetime, this means we define an embedding map
\begin{equation} \label{2.65}
    \phi: {\cal N}\to{\cal M}\,,\qquad \phi:x^{a}\to(\Omega_{\cN}(x^a),x^{a})\,.
\end{equation}
Then, $\cal N$ is located at the points in spacetime where $\Omega=\Omega_{\cal N}(x^{a})$. As such, the normal one-form to $\cal N$ can be defined as
\begin{equation} \label{nOmega}
    n = n_{\mu}dx^{\mu} = d(\Omega - \Omega_{\cal N})\,.  
\end{equation}
The dual vector is
\begin{equation} \label{2.67}
    n^{\mu}\pa_{\mu} = e^{-2\beta}\left(2F\Omega^{2}-(\pa_{u}+U^{A}\pa_{A})\Omega_{\cal N}\right)\pa_{\Omega} + e^{-2\beta}\pa_{u} + \left(e^{-2\beta}U^{A}-\hat{g}^{AB}\pa_{B}\Omega_{\cal N}\right)\pa_{A}\,.
\end{equation}
Then, one can compute the norm  of $n$ evaluated at $\Omega=\Omega_{\cal N}(x^{a})$,
\begin{equation} \label{2.68}
    |n|^{2} = e^{-2\beta}\left(2F\Omega_{\cal N}^{2}-(\pa_{u}+U^{A}\pa_{A})\Omega_{\cal N}\right) - e^{-2\beta}\pa_{u}{\Omega_{\cal N}} - \left(e^{-2\beta}U^{A}-\hat{g}^{AB}\pa_{B}\Omega_{\cal N}\right)\pa_{A}\Omega_{\cal N}\,.
\end{equation}

In order to foliate the conformal spacetime by a family of null hypersurfaces, we must solve for $|n|^2=0$ perturbatively as we push $\cal N$ to the boundary $\mathscr{I}^{+}$. We follow a similar construction to that in \cref{section:geometric setup} and introduce the following perturbative expansion for $\Omega_{\cal N}$
\begin{equation} \label{2.69}
    \Omega_{\cal N}(x^{a}) = \tilde{c}_{1}(x^a)\lambda + \tilde{c}_{2}(x^a)\lambda^{2} + \tilde{c}_{3}(x^a)\lambda^{3} + {\cal O}(\lambda^{4})\,.
\end{equation}
The conformal boundary is now reached in the limit $\lambda\to 0$. Plugging the expansion~\eqref{2.69} into~\eqref{2.68} and perturbatively setting $|n|^{2}=0$ up to ${\cal O}(\lambda^{3})$ yields the following constraint equations for the coefficients
\begin{equation} \label{2.70}
    \partial_{u}\tilde{c}_{1} = 0\,, \qquad \partial_{u}\tilde{c}_{2} =  \frac{1}{2}\gamma^{AB}\partial_{A}\tilde{c}_{1}\partial_{B}\tilde{c}_{1} + \tilde{c}_{1}^{2}\bF\,,
\end{equation}
and
\begin{equation} \label{2.71}
    \pa_u \tilde{c}_3=2\bF  \tilde{c}_1 \tilde{c}_2+\gamma^{AB}\pa_A \tilde{c}_1\pa_B \tilde{c}_2-\frac12 \tilde{c}_1 C^{AB}\pa_A \tilde{c}_1\pa_B \tilde{c}_1+2\bbeta \tilde{c}_1^2\pa_u \tilde{c}_1- \tilde{c}_1^2\bU ^A\pa_A \tilde{c}_1-\tilde{c}_1^3 m\,.
\end{equation}
For simplicity, we will from now on set $\tilde{c}_{1}=1$. Then, the first condition in~\eqref{2.70} is trivially satisfied. The remaining two conditions simply become 
\begin{equation} \label{conformalNullConstraints}
    \pa_{u}\tilde{c}_{2} = \bF\,,  \qquad \pa_{u}\tilde{c}_{3} = 2\bF \tilde{c}_{2}-m\,.
\end{equation}
We are seeing here the same physics as we saw in the physical spacetime, namely, that the curvature of a cut ${\cal C}$ and the Bondi mass aspect are contributing in such a way as to keep the surface null as we push it to ${\mathscr{I}}^{+}$.\footnote{The comparison with the physical spacetime becomes explicit upon noting that, with $\lambda_{conf}=1/\lambda_{phys}$, one has $\tilde{c}_2=-c_0$ and $\tilde{c}_3=c_0^2-c_1$.} For unit round-sphere cuts ${\cal C}$, $\bF=1/2$, and we can consistently take $\tilde{c}_{2}$ to be independent of the angular coordinates, $\pa_{A}\tilde{c}_{2}=0$.

\subsubsection*{Null Rigging and Intrinsic Carrollian Data}
\label{IntrinsicCarrollianData}

In order to set up the rigging construction, we choose the following rigging vector field
\begin{equation} \label{2.73}
    k^\mu\pa_\mu=-\partial_\Omega\,,
\end{equation}
which satisfies
\begin{equation} \label{2.74}
    |k|^2=0\,,\qquad k^\mu n_\mu=-1\,.
\end{equation}
The associated one-form is
\begin{equation} \label{2.75}
    k_\mu dx^\mu=-e^{2\beta}du\,.
\end{equation}
We can again use the normal $n$ and rigging vector $k$ to define the projector onto ${\cal N}$,
\begin{align}
    \Pi_\mu{}^\nu=\delta_\mu^\nu+n_\mu k^\nu\,,
\end{align}
which induces the intrinsic ruled Carrollian structure as in~\eqref{carrollvec}. 

We want to expand the intrinsic Carrollian data, as well as $\kappa$ and $\pi_a$, in powers of $\lambda$ as before. Here, the boundary is reached as $\lambda\to0$. The Carrollian vector field admits the following expansion
\begin{align}
    \ell^{u} &=1-2\bbeta\lambda^{2} + {\cal{O}} (\lambda^{3})\,,  \\
    \ell^{A} &= \lambda^{2}\bU^{A} + \lambda^{3}\overset{(3)}{\ell}\vphantom{\ell}^{A} + {\cal O}(\lambda^{4})\,,
\end{align}
where
\begin{align}
    \overset{(3)}{\ell}\vphantom{\ell}^{A} = \left(U_{1}^{A} + 2\tilde{c}_{2}\bU^{A} - \pa^{A}\tilde{c}_{3}\right)\,.
\end{align}
Next, for the induced Carrollian metric we have 
\begin{align} 
    q_{uu}	&=\mathcal{O}(\lambda^{4})\,,\\
    q_{uA}&=-\lambda^{2}\bU_{A}-\lambda^{3}\left(C_{AB}\bU^{B}+2\tilde{c}_{2}\bU_{A}+U_{1 A}-\partial_{A}\tilde{c}_{3}\right)+\mathcal{O}(\lambda^{4})\,,\\
    q_{AB}	&=\gamma_{AB}+\lambda C_{AB}+\lambda^{2}\left(\tilde{c}_{2}C_{AB}+\frac{1}{4}\gamma_{AB}C_{CD}C^{CD}\right)+\mathcal{O}(\lambda^{3})\,.\label{2.86}
\end{align} 
And finally, the Ehresmann connection,
\begin{align}
    k_u=-1-2\bbeta\lambda^2+\mathcal O(\lambda^3)\,,\qquad k_A=0\,.
\end{align}
As for the inaffinity and Hájíček one-form, we have
\begin{equation} \label{conformalkappa} 
    \kappa=2\bF \lambda+\left(2\tilde{c}_{2}\bF -3m\right)\lambda^{2}+\mathcal{O}(\lambda^{3})\,,
\end{equation}
and 
\beq 
   \pi_{u}	&=\mathcal{O}(\lambda^{3})\, , \qquad
    \pi_{A}&=\lambda\bU_{A}-\lambda^{2}\left(\bP_{A}+2\partial_{A}\bbeta-\tilde{c}_{2}\bU_{A}\right)+\mathcal{O}(\lambda^{3}) \,,\label{AconformalHaj}
\eeq
respectively.

Paralleling our physical spacetime analysis, we next evaluate the components of the expansion tensor. In the asymptotic limit, they are given by
\begin{align}
    \theta_{uu}	&=\mathcal{O}(\lambda^{4})\, , \\
    \theta_{uA}	&=-\frac{\lambda^{3}}{2}N_{AB}\bU^{B}+\mathcal{O}(\lambda^{4}) \, ,\\
    \theta_{AB}	&= \frac{\lambda}{2}N_{AB} + \frac{\lambda^{2}}{2}\left(\bF C_{AB}+\tilde{c}_{2}N_{AB} + \frac{1}{2}\gamma_{AB}N_{CD}C^{CD} + \mathcal{L}_{\bU}\gamma_{AB}\right) + \mathcal{O}(\lambda^{3})\,.
\end{align}
We can raise an index, as explained in \cref{asym}. This yields asymptotic expansions for the components of $\theta_{a}{}^{b}$
\begin{align}
    \theta_{a}{}^{u}&=0\,, \\
        \theta_{u}{}^{A}&=-\frac{\lambda^{3}}{2}N^{AB}\bU_{B}+\mathcal{O}(\lambda^{4})\,,\\
    \theta_{A}{}^{B}	
    &=\frac{\lambda}{2}N_{A}{}^{B} +\frac{\lambda^{2}}{2}\left(\bF C_{A}{}^{B}+\tilde{c}_{2}N_{A}{}^{B}-N_{AC}C^{BC} +\frac{1}{2}\delta_{A}^{B}C^{CD}N_{CD} +\gamma^{BC}{\cal L}_{\bU}\gamma_{AC}\right) + {\cal O}(\lambda^{3})\,. 
\end{align}
Taking a trace of this, we find the expansion scalar
\begin{align} \label{conformaltheta}
    \theta =\lambda^{2}\bnabla _{A}\bU^{A} + {\cal O}(\lambda^{3})\,.
\end{align}
Similar manipulations allow one to compute the shear, resulting in 
\begin{align}
    \sigma_{a}{}^{u}	&=0\,, \label{conformalShear_a^u}\\
    \sigma_{u}{}^{A}&=-\frac{\lambda^{3}}{2}N^{AB}\bU_{B}+\mathcal{O}(\lambda^{4})\,,  \label{conformalShear_u^A} \\
    \sigma_{A}{}^{B}	&=\frac{\lambda}{2}N_{A}{}^{B}\notag\\
    &\quad+\frac{\lambda^{2}}{2}\left(\bF C_{A}{}^{B} + \tilde{c}_{2}N_{A}{}^{B} - N_{AC}C^{BC} + \frac{1}{2}\delta_{A}{}^{B}N^{CD}C_{CD} + \mathcal{L}_{\bU }\gamma_{\langle A}{}^{B\rangle}\right) + {\cal O}(\lambda^{3})\,.
\label{conformalShear_A^B}
\end{align}
We next use these expansions to study the asymptotic limit of the intrinsic gravitational constraints along the null foliation in the conformally compactified spacetime.

\subsection{Asymptotic Einstein Equations}

We now study the asymptotic limit of the projected Einstein equations in the conformally compactified spacetime. We will recover the BMLF and angular momentum equation at ${\cal O}(\lambda^{2})$. Just as in the physical spacetime, our starting point is first the Raychaudhuri constraint
\begin{equation} \label{ConfRay}
    C = ({\cal L}_{\ell}+\theta)\theta - \mu\theta + \sigma_{a}{}^{b}\sigma_{b}{}^{a} + \ell^{a}\ell^{b}\hat{R}_{ab}\,.
\end{equation}
The main difference in the conformally compactified spacetime is that the Ricci tensor is non-vanishing on shell, due to its non-trivial transformation under conformal rescalings of the metric. Hence, the last term in~\eqref{ConfRay} needs to be treated with care. Using the rigging projector, this term can be written in terms of bulk spacetime indices
\begin{align} \label{3.51}
    \ell^{a}\ell^{b}\hat{R}_{ab}\overset{\cal N}{=}n^{\mu}n^{\nu}\hat{R}_{\mu\nu}\,.
\end{align}
Using our nullity constraints in~\eqref{conformalNullConstraints} and expanding asymptotically, the leading order of~\eqref{3.51} becomes  
\begin{align}
    \ell^{a}\ell^{b}\hat{R}_{ab} = 2\lambda^{2}\pa_{u}m + {\cal O}(\lambda^{3})\,.
\end{align}
Next, from $\kappa$ in~\eqref{conformalkappa} and $\theta$ in~\eqref{conformaltheta}, it is easy to see that the $\mu\theta$ term in~\eqref{ConfRay} is ${\cal O}(\lambda^{3})$, and can thus be neglected. Then, looking at the asymptotic expansion of the components of the shear in~\eqref{conformalShear_A^B}, we have 
\begin{equation} \label{shear_squared_conformal}
    \sigma_{a}{}^{b}\sigma_{b}{}^{a} = \frac{\lambda^{2}}{4}N_{AB}N^{AB} + {\cal O}(\lambda^{3})\,.
\end{equation}
Finally, to leading order, the first term of~\eqref{ConfRay} is simply
\begin{align}
    \mathcal{L}_{\ell}\theta+\theta^{2} &= \pa_{u}\theta + {\cal O}(\lambda^{4}) \notag \\
    &=-\frac{\lambda^{2}}{2}\bnabla _{A}\bnabla_{B}N^{AB} + {\cal O}(\lambda^{3})\,, 
\end{align}
where we used our expression for $\theta$ in~\eqref{conformaltheta}. Putting all these terms together, we arrive at the following asymptotic limit of the Raychaudhuri constraint 
\begin{align}
    C = \lambda^{2}\left(-\frac{1}{2}\bnabla _{A}\bnabla _{B}N^{AB} + \frac{1}{4}N_{AB}N^{AB} + 2\pa_{u}m\right) + {\cal O}(\lambda^{3})\,.
\end{align}
Going on-shell we arrive at 
\begin{align}
    \pa_{u}m = \frac{1}{4}\bnabla _{A}\bnabla _{B}N^{AB} - \frac{1}{8}N_{AB}N^{AB}\,,
\end{align}
which is the BMLF. Although, as expected, the final equation agrees with the physical spacetime result \eqref{BMLF}, the individual contributions
to the Raychaudhuri constraint are organized differently. In the physical description, the Bondi mass aspect enters through the subleading expansion
of the null generators. In the conformally
compactified description, instead, its contribution arises from the Ricci tensor, which is non-vanishing even though the physical spacetime
is Ricci-flat. Thus, the agreement between the two descriptions is not term by term: the conformal transformation redistributes the contributions to
the intrinsic gravitational constraints while preserving their asymptotic dynamics.

Turning now to the Damour constraint
\begin{align}
    C_{a} &= (\mathcal{L}_{\ell}+\theta)\pi_{a}+\theta\varphi_{a} - (\bar{D}_{b}+\varphi_{b})(\mu q_{a}{}^{b} - \sigma_{a}{}^{b})- q_a{}^b \hat{R}_{bc}\ell^c\,,\label{ConfCompactDconstraint}
\end{align}
we first expand the Carrollian acceleration
\begin{align}
    \varphi_{u}={\cal O}(\lambda^{4})\,, \qquad \varphi_{A}=2\lambda^{2}\pa_{A}\bbeta + {\cal O}(\lambda^{3})\,.
\end{align}
As in the physical spacetime, terms in the Damour constraint that involve $\varphi$ will be subleading since $\varphi_{a}$ multiplies quantities that go like ${\cal O}(\lambda)$ at best. With this, the temporal component of the Damour constraint is subleading
\begin{align}
    C_{u} = {\cal O}(\lambda^{3})\,.
\end{align}

As for the spatial components $C_{A}$, let us first consider the conformal Ricci tensor term. Since $q_{a}{}^{u}=0$ and $q_{A}{}^{B}=\delta_{A}{}^{B}$, we have
\begin{align}
    q_{A}{}^{b}\hat{R}_{bc}\ell^{c} = \hat{R}_{Ab}\ell^{b}\,.
\end{align}
Now, using its definition in~\eqref{hatRicciTensor}, as well as going onto the surface $\Omega=\Omega_{\cal N}$, we obtain
\begin{align}
    q_A{}^b\hat{R}_{bc}\ell^c = 2\lambda^2\left(\partial_A m + \bF\bU_A\right)+{\cal O}({\lambda}^3)\,.
\end{align}

Next, the first term in~\eqref{ConfCompactDconstraint} is expanded as
\begin{align}
    ({\cal L}_{\ell} + \theta)\pi_{A} = \pa_{u}\pi_{A} + {\cal O}(\lambda^{3})\,,
\end{align}
where we have used that $\ell^{u}\sim{\cal O}(1)$, $\ell^{A}\sim{\cal O}(\lambda^{2})$, $\theta\sim{\cal O}(\lambda^{2})$ and $\pi_{A}\sim{\cal O}(\lambda)$. Plugging in our expansion for the Hájíček connection, this term becomes
\begin{align}
    ({\cal L}_{\ell} + \theta)\pi_{A} = \lambda\pa_{u}\bU_{A} +\lambda^{2}\left( \frac{1}{8} \bnabla_{A}\left(N^{BC}C_{BC}\right)+\pa_{u}(\tilde{c}_{2}\bU_{A})-\pa_{u}\bP_{A}\right) + {\cal O}(\lambda^{3})\,, 
\end{align}
where we have used the on-shell value for $\bar\beta$ in~\eqref{2.6}, which still holds in the conformally compactified spacetime. Finally, we have the term
\begin{align}
    \bar{D}_{b}(\mu q_{A}{}^{b} - \sigma_{A}{}^{b}) &= q_{b}{}^{c}q_{A}{}^{d}q_{e}{}^{b}D_{c}(\mu q_{d}{}^{e}-\sigma_{d}{}^{e}) \\
    &= \pa_{A}\mu - q_{A}{}^{d}q_{e}{}^{c}D_{c}\sigma_{d}{}^{e}\,.
\end{align}
Using the expansions for $\kappa$~\eqref{conformalkappa} and $\theta$~\eqref{conformaltheta}, the first term can be expanded as
\begin{align}
    \pa_{A}\mu = \lambda^{2}\left(\frac{1}{2}\bnabla_{A}\bnabla_{B}\bU^{B}-3\pa_{A}m\right) + {\cal O}(\lambda^{3})\,.
\end{align}
As for the second term, we have
\begin{align}
    q_{A}{}^{d}q_{e}{}^{c}D_{c}\sigma_{d}{}^{e} &= q_{A}{}^{d}q_{e}{}^{c}\left(\pa_{c}\sigma_{d}{}^{e} + \Gamma^{e}_{cf}\sigma_{d}{}^{f}-\Gamma^{f}_{cd}\sigma_{f}{}^{e}\right) \\
    &= q_{A}{}^{d}q_{e}{}^{c}\left(\pa_{c}\sigma_{d}{}^{e} + \chi_{cf}\ell^{e}\sigma_{d}{}^{f} + \overset{q}{\Gamma}\vphantom{\Gamma}^{e}_{cf}\sigma_{d}{}^{f} - \chi_{cd}\ell^{f}\sigma_{f}{}^{e} - \overset{q}{\Gamma}\vphantom{\Gamma}^{f}_{cd}\sigma_{f}{}^{e}\right)\,,
\end{align}
where we have defined
\beq
    \Gamma^a_{bc}=\chi_{bc}\ell^a+\overset{q}{\Gamma}\vphantom{\Gamma}^{a}_{bc}\,,
\eeq
with
\beq \label{qChristoffel}
    \overset{q}{\Gamma}\vphantom{\Gamma}^{a}_{bc}=\frac{q^{ad}}{2}(\pa_b q_{cd}+\pa_c q_{bd}-\pa_d q_{bc})\,.
\eeq
Using that $\ell^{e}q_{e}{}^{c}=0$ and $\ell^{f}\sigma_{f}{}^{e}=0$, the terms involving $\chi_{ab}$ vanish. In addition, from $q^{au}=0$, $\sigma_{a}{}^{u}=0$, we have
\begin{align}
    q_{A}{}^{d}q_{e}{}^{c}D_{c}\sigma_{d}{}^{e} = \overset{q}{D}_{B}\sigma_{A}{}^{B}\,,
\end{align}
where $ \overset{q}{D}_B\sigma_A{}^B=\pa_B\sigma_A{}^B-\overset{q}{\Gamma}\vphantom{\Gamma}^{C}_{BA}\sigma_C{}^B+\overset{q}{\Gamma}\vphantom{\Gamma}^{B}_{BC}\sigma_A{}^C$, with
\begin{align}
    \overset{q}{\Gamma}\vphantom{\Gamma}^{C}_{AB} = \frac{q^{CD}}{2}(\pa_{A}q_{DB} + \pa_{B}q_{AD}-\pa_{D}q_{AB})\,.
\end{align}
Using the expansions of $q_{AB}$ and $\sigma_{A}{}^{B}$, we arrive at
\begin{align}
    \overset{q}{D}_{B}\sigma_{A}{}^{B} = \frac{\lambda}{2}\bnabla_{B}N_{A}{}^{B}+\lambda^{2}\left( \bnabla _{B}\overset{(2)}{\sigma}_{A}{}^{B}- \frac{1}{4}N^{BC}\bnabla_{A}C_{BC}\right) + {\cal O}(\lambda^{3})\,,
\end{align}
where 
\begin{align}
    \overset{(2)}{\sigma}_{A}{}^{B} = \frac{1}{2}\left(\bF C_{A}{}^{B} + \tilde{c}_{2}N_{A}{}^{B} - N_{AC}C^{BC} + \frac{1}{2}\delta_{A}^{B}N^{CD}C_{CD} + \mathcal{L}_{\bU }\gamma_{\langle A}{}^{B\rangle}\right)\,, 
\end{align}
is the subleading piece of the shear in~\eqref{conformalShear_A^B}.

Summing up all the terms in the spatial part of the Damour constraint, at ${\cal O}(\lambda^{2})$ we have
\begin{align} \label{3.73}
    C_{A} = \lambda^{2}\left(-\pa_{u}\bP_{A}+\frac{1}{8}\bnabla _{A}\left(N^{BC}C_{BC}\right) + \pa_{u}(\tilde{c}_{2}\bU _{A})-\frac{1}{2}\bnabla _{A}\bnabla _{B}\bU ^{B}+\pa_{A}m\right. \notag \\
    \left.-2\bF\bU_A -\frac{1}{4}N^{BC}\bnabla _{A}C_{BC} + \bnabla _{B}\overset{(2)}{\sigma}_{A}{}^{B}\right) + {\cal O}(\lambda^{3})\,.
\end{align}
Further processing the last term in~\eqref{3.73}, we have
\begin{align}
    \bnabla _{B}\overset{(2)}{\sigma}_{A}{}^{B} = -\pa_{u}(\tilde{c}_{2}\bU _{A}) + \frac{1}{4}\bnabla _{B}\left(C_{AC}N^{BC}-N_{AC}C^{BC}\right) + \frac{1}{2}\bnabla _{C}\bnabla ^{C}\bU _{A} + \bF \bU _{A}\,,
\end{align}
where we have again made use of the identity $\delta_{A}{}^{B}N_{CD}C^{CD}=C_{AC}N^{BC}+N_{AC}C^{BC}$. 

Plugging this into the above limit and using similar manipulations as those we employed for the Damour equation in the physical spacetime we arrive at
\begin{align}
        C_{A} &= \lambda^{2}\Big(-\pa_{u}\bP_{A}+\pa_{A}m+\frac{1}{8}\bnabla _{A}(N^{BC}C_{BC}) -\frac{1}{4}N^{BC}\bnabla _{A}C_{BC}\nonumber\\
        &\quad+ \frac{1}{4}\bnabla _{B}(C_{AC}N^{BC}-N_{AC}C^{BC})+ \frac{1}{2}\bnabla _{C}\big(\bnabla ^{C}\bU _{A} - \bnabla _{A}\bU ^{C}\big)\Big) + {\cal O}(\lambda^{3})\,.
\end{align}
Setting $C_A=0$ and using $\bU^A=-\frac12\bnabla_BC^{AB}$ reproduces the angular momentum equation obtained in~\cref{sec3}. This closes the analysis of the asymptotic limit of the geometric, kinematical, and dynamical quantities on a family of null hypersurfaces. 

Having recovered the Bondi evolution equations in the conformally compactified description, we next turn to the asymptotic limit of the gravitational symplectic potentials.

\subsection{Asymptotic Symplectic Potentials}

We now express the physical Einstein--Hilbert symplectic potential in terms of the conformal metric and determine its asymptotic limit. We impose the same phase-space restrictions as in the physical analysis, namely $\delta\gamma_{AB}=0$, $\delta\tilde c_2=0$ and $\delta\lambda=0$, and keep the conformal factor fixed under variations, $\delta\Omega=0$. Starting from~\eqref{EHPot} and substituting $g_{\mu\nu}=\Omega^{-2}\hat g_{\mu\nu}$, we obtain
\begin{align} \label{4.16}
    \Theta^\mu(\hat g)
    =\frac{1}{\Omega^2}\theta_{\rm EH}^\mu(\hat g)+\hat V^\mu\,.
\end{align}
with
\begin{align}
    \hat{V}^{\mu} = \frac{1}{\Omega^{3}}\pa_{\alpha}\Omega\left[\hat{g}^{\mu\alpha}\hat{g}^{\lambda\sigma}-4\hat{g}^{\mu\lambda}\hat{g}^{\alpha\sigma}\right]\delta\hat{g}_{\lambda\sigma}\,.
\end{align}  
We retain the spacetime orientation fixed in the physical analysis, so that $\hat\varepsilon_4=\Omega^4\varepsilon_4$.

Writing~\eqref{4.16} as a spacetime three-form and pulling it back to ${\cal N}$ we obtain
\begin{align} \label{4.21}
    \underset{\leftarrow}{\Theta}=
    \frac{1}{\Omega_{\cal N}^2}\theta_{\rm EH}(\hat g)
    -\frac{\hat\varepsilon_{\cal N}}{16\pi G}n_\mu\hat V^\mu\,.
\end{align}
Here, $\hat\varepsilon_{\cal N}:=i_k\hat\varepsilon_4$, with the rigging chosen in~\eqref{2.73}. We can then conformally transform the $(b,c)$ representative as it appears in~\eqref{EHtheta}, obtaining
\begin{align}
    \Theta^{(b,c)}=\Theta + \delta\hat\ell^{(b,c)} -d\hat\vartheta^{\rm EH}\,,
\end{align}
where the hatted quantities are the conformally transformed boundary Lagrangian and corner potential, respectively.

Let us analyze the asymptotic limit of $\Theta^{(b,c)}$, by taking the limit of the three pieces composing it. First, one has
\beq
    \lim_{\lambda\to 0} \Theta=\theta^\mathrm{AS}+\frac{\varepsilon_\mathcal N^{(0)}}{16\pi G}\delta\left(2m-\frac{1}{2}\bnabla_A\bnabla_BC^{AB}-\frac{1}{8}C^{AB}N_{AB}\right)\,.
\eeq
Secondly, we get
\beq
    \lim_{\lambda\to 0} \delta\hat\ell^{(b,c)}=\frac{\varepsilon_\mathcal N^{(0)}}{16\pi G}\delta\left((b-2c) m-c \bnabla_A\bU^A\right)\,.
\eeq
Eventually
\beq
    \lim_{\lambda\to 0} d\hat\vartheta^{\rm EH}=\frac{\varepsilon_\mathcal N^{(0)}}{16\pi G}\left[-\frac{1}{8}\delta\left(C^{AB}N_{AB}\right)+\frac{1}{2}\bnabla_A\bnabla_B\delta C^{AB}\right]\,.
\eeq

Putting things together, we thus obtain 
\begin{align} \label{4.34}
    \lim_{\lambda\to 0}\Theta^{(b,c)} = \theta^\mathrm{AS}+\frac{\varepsilon_\mathcal N^{(0)}}{16\pi G}\delta\left[(b-2c+2)m+\frac{c-2}{2}\bnabla_A\bnabla_BC^{AB}\right]\,.
\end{align}
This equation agrees exactly with the physical spacetime result \eqref{physical bc limit}. Once again, however,
this agreement is not realized term by term. The conformal transformation redistributes the finite contributions among the Einstein--Hilbert
potential, the boundary Lagrangian and the corner term. In particular, the conformal contribution
appearing in \eqref{4.21} has the finite asymptotic limit
\begin{align} \label{1.26}
    \lim_{\lambda\to0}n_{\mu}\hat{V}^{\mu}\hat\varepsilon_{\cal N} = -8\delta m\,\ve^{(0)}_{\cN}\,,
\end{align}
and therefore introduces a contribution proportional
to the variation of the Bondi mass aspect. In~\eqref{4.21}, this contribution combines with the rescaled conformal Einstein--Hilbert potential to reproduce the physical Einstein--Hilbert potential at null infinity.

We thus find that, within the boundary conditions
considered here, the physical and conformally
compactified descriptions reproduce the same
asymptotic gravitational constraints and the same
family of symplectic potentials. The difference lies
in how the individual geometric and symplectic
contributions combine to produce these results,
rather than in the resulting asymptotic structures.
\end{appendix}

\bibliographystyle{uiuchept}
\bibliography{Arxiv-V1}

\end{document}